\documentclass[aps,showpacs,preprintnumbers,amsmath,amssymb,nofootinbib]{revtex4}

\usepackage{graphicx}
\usepackage{color}

\def \be  {\begin{equation}}
\def \ee  {\end{equation}}
\def \ee  {\end{equation}}
\def \bea {\begin{eqnarray}}
\def \eea {\end{eqnarray}}

\newcommand{\nn}{\nonumber}
\newcommand{\de}{\partial}

\begin{document}

\preprint{ECTP-2015-01}
\preprint{WLCAPP-2015-01}

\title{Review on Generalized Uncertainty Principle}
\author{Abdel Nasser TAWFIK}
\email{a.tawfik@eng.mti.edu.eg}
\affiliation{Egyptian Center for Theoretical Physics (ECTP), Modern University for Technology and Information (MTI), 11571 Cairo, Egypt}
\affiliation{World Laboratory for Cosmology And Particle Physics (WLCAPP), 11571 Cairo, Egypt}

\author{Abdel~Magied~DIAB}
\affiliation{Egyptian Center for Theoretical Physics (ECTP), Modern University for Technology and Information (MTI), 11571 Cairo, Egypt}
\affiliation{World Laboratory for Cosmology And Particle Physics (WLCAPP), 11571 Cairo, Egypt}

\begin{abstract}
Based on string theory, black hole physics, doubly special relativity and some "thought" experiments, minimal distance and/or maximum momentum are proposed. As alternatives to the generalized uncertainty principle (GUP), the modified dispersion relation, the space noncommutativity, the Lorentz invariance violation, and the quantum-gravity-induced birefringence effects are summarized. 
The origin of minimal measurable quantities and the different GUP approaches are reviewed and the corresponding observations are analysed. Bounds on the GUP parameter are discussed and implemented in understanding recent PLANCK observations on the cosmic inflation. The higher-order GUP approaches predict minimal length uncertainty with and without maximum momenta. 

\end{abstract}


\maketitle

\section{Introduction}
\label{intro}

Deduced from different approaches related to the quantum gravity (QG) such as black hole physics \cite{3,4} and string theory \cite{guppapers,5}, a minimal length was predicted \cite{Tawfik:2014zca}. The fundamental idea is simple. The string is conjectured not to interact at distances smaller than its size, which is determined by its tension. Information about the string interactions would be included in the Polyakov loop action \cite{daxson}. The existence of a minimal length leads to generalized (Heisenberg) uncertainty principle (GUP) \cite{guppapers}. At Planck (energy) scale, the corresponding Schwarzschild radius becomes comparable to the Compton wavelength. In presence of gravitational effects, the higher energies (Planck energy) result in further decrease in the Schwarzschild radius $\Delta x$ and therefore, $\Delta x \approx \ell_{Pl}^2 \Delta p/\hbar$. From this observation and the ones deduced from various {\it gedanken} experiments \cite{Tawfik:2014zca}, it was suggested that the GUP approaches would  be very essential, especially at some concrete scales of energies and distances.

According to the Heisenberg uncertainty principle (HUP), which represents one of the fundamental properties of quantum systems, there should be a fundamental limit for the measurement accuracy, with which certain pairs of physical observables, such as position and momentum and energy and time, can not be measured, simultaneously. In other words, the more precisely one quantity is measured, the less precise the other one shall be detected. In quantum mechanics (QM), the physical observables are described by operators in Hilbert space. Given an observable $A$, an operator is defined as a standard deviation of $A$, $\Delta A=A-\langle A \rangle$, where its expectation value reads $\langle (\Delta A)^2 \rangle = \langle {A^2} \rangle - {\langle A \rangle}^2$.
Using Schwartz inequality \cite{Schwarz} $\langle \alpha | \alpha\rangle \langle \beta | \beta \rangle \geq | \langle \alpha | \beta \rangle |^2$, 
which is valid for any ket- and bra-state $|\alpha\rangle =\Delta A | \alpha^{'} \rangle$ and $|\beta \rangle = \Delta B | \beta^{'} \rangle$, respectively. In Dirac algebra, Cauchy-Schwartz inequality implies that $(\Delta A)^{2}\; (\Delta B)^{2}\geq \frac{1}{4} | \langle \Delta A\; \Delta B \rangle |^{2}$,  
\bea
\Delta\, A\,  \Delta\, B &\geq & \frac{1}{2} | \langle \Delta\, A\; \Delta\, B \rangle |. 
\eea

In Heisenberg algebra, the position  and momentum operator, $\hat{x}$ and $\hat{p}$, respectively, satisfy the canonical commutation relation $[\hat{x} ,\hat{p}]= \hat{x} \hat{p} -\hat{p} \hat{x}= i \hbar$.  As a consequence, their measurement uncertainties, $\Delta x $ and $\Delta p$, respectively, (Heisenberg uncertainty principle) are related to each other (in natural units)
\be 
\Delta x\; \Delta p \geq \frac{\hbar}{2}.
\ee 
  
In detecting an arbitrarily small length scale, one has to utilize tools of sufficiently high energy (high momentum) and thus very short wavelength. This is noting but the the principle of the high-energy colliders/accelerators. But, there are reasons to believe that at high energies, the gravity becomes dominant. Thus, the linear indirect relation between energy and wavelength should be violated.

The detectability of quantum space-time foam with gravitational wave interferometers was discussed \cite{disct1} and  criticized \cite{disct1,disct2} due to the limited measurability of the smallest quantum distances. Furthermore, Wigner inequalities \cite{wigner57,wigner58} were implemented in describing the quantum constraints on the black-hole lifetime \cite{barrow96}, which is proportional to the Hawking lifetime. The latter is to be calculated under the assumption that the black hole is a black body and therefore it follow Stefan-Boltzmann law. It was found that the Schwarzschild radius is correspondent to the constraints on Wigner size and the power of information processing in black hole can be estimated through the emission of Hawking radiation \cite{swLitr}. 

The relationship between the minimal length and maximum momentum is  presented. As introduced in previous sections, there are various approaches to GUP proposing the existence of nonvanishing minimal length that leads to non-commutative geometry.

Recently, it was proposed that GUP would also arise naturally in the horizon wave-function formalism, which is obtained from modelling the electrically charged source in the inner Cauchy horizon of Reissner-Nordstr¨om black hole as a Gaussian wave-function \cite{hwf1,hwf2}. Significant ranges for the black hole mass and the specific charge were found, for which the probability of realising the inner horizon becomes negligible. The latter suggests the existence of a minimum black hole mass and eventually a minimum charge, and that any semiclassical instability expected near the inner horizon may not occur in quantum black holes.

The present review article is organized as follows. The modified dispersion relation are introduced in section \ref{sec:mdr}.
The space noncommutativity is reviewed in section \ref{sec:nc}. Section \ref{sec:liv} discusses the Lorentz invariance violation and their experimental tests. The quantum-gravity-induced birefringence effect is elaborated in section \ref{sec:bire}.
The origin of minimal measurable quantities shall be summarized in section \ref{sec:orgn}.

In section \ref{sec:gup1}, we summarize the behavior of some well-known expressions for GUP. These expressions contain quadratic term of momenta with a minimal uncertainty on position. In section \ref{ssec:1}, we shall investigate the modification of the uncertainty relation due to the high-energy fixed-angle scatterings at short length such as the string length. In section \ref{ssec:2}, the uncertainty relation through various {\it gedanken} experiments which are designed to measure the area of the apparent horizon of black hole is reviewed. These thought experiments assume QG due to recording the photons of the Hawking radiation, which are emitted from the apparent horizon. Due to quantized space-time of the QFT and the geometric approach to curvature of momentum space, an algebraic approach can be expressed in the coproducts and the description of the Hopf-algebra \cite{MR28} leading to modified commutation relation between position and momenta, section \ref{ssec:3}. In section \ref{ssec:5}, a new commutation relation containing a linear term as an addition of the quadratic term of momenta and predicts of the maximum measurable of momenta, shall be investigated.

In section \ref{sec:min1}, the relations describing the minimal length uncertainty are outlined. Two proposals for the modification of the momentum operator are introduced. The proposal of a minimal length uncertainty with a further modification in the momentum shall be reviewed. The main features in Hilbert space representation of QM of the minimal length uncertainty will be studied. Furthermore, their difficulties are also listed out. We show how to overcome these difficulties, especially  in Hilbert space representation.

In section \ref{sec:max}, the GUP approaches relating to string theory and black hole physics (lead to a minimum length) and the ones relating to DSR (suggest a similar modification of commutators) shall be studied. The main features and difficulties in Hilbert space representation will be reviewed, as well, and we show how to overcome these difficulties.

In section \ref{other}, other alternative approaches to GUP such as the one suggested by Nouicer \cite{Nouicer}, in which an exponential term of momentum and minimal length appears, shall be introduced. This approach agrees well with the GUP which is originated in the theories for QG. There is another approach coming up with higher orders of  the minimal length uncertainty and maximal observable momentum. Finally, we compare between these approaches.
 
Before introducing GUP approaches, we review the possibilities of modifying the dispersion relations, section \ref{sec:mdr}, and space noncommutativity, section \ref{sec:nc}.

\section{Modified dispersion relations}
\label{sec:mdr}

There are various experimental measurements indicating that the Lorentz invariance principle would be violated, at high energies \cite{Tawfik:2014zca,Tawfik:2012hz}. The velocity of light $c$ is conjectured to differ from the maximum attainable velocity of a material body. Such a small adjustment of $c$ leads to modification of the energy-momentum relation and to add possible $\delta \textit{v}$ \cite{Glashow:1a,Glashow:1b,Glashow:2,Glashow:3,Amelino98} to the dispersion relation in vacuum state, which could be sensitive to a type of candidate QG effect, that has been recently considered in particle physics literature. In additional to that, the possibility that the relation connecting energy and momentum in special relativity may be modified at Planck scale because of the threshold anomalies of ultra-high energy cosmic ray (UHECR) is conventionally named as modified dispersion relations (MDRs) \cite{Amelino98,Amelino2001,Amelino2001b,Amelino2004,Amelino2004b,Amelino2006,Nozari2006,Aloisio,Jacobson,Nozari}. Successful searches would discover a connection between the particle physics and cosmology \cite{Glashow:1a,Glashow:1b,Glashow:2,Glashow:3}. The speed of light is not limited to that. Various research works are devoted to studying the modification of the energy-momentum conservations laws of interactions, such as pion photo-production through the inelastic collisions of the cosmic-ray nucleons with cosmic microwave background, and the higher energy photon propagating in the intergalactic medium which would suffer from inelastic impacts with photons in the infrared background and results in an enhanced production of electron-positron pair. \cite{Y. J. Ng:1995,Y. J. Ng:2000}.

\subsection{Modified dispersion relation and generalized uncertainty principle}
Many researches of loop-quantum-gravity studies\cite{lqgDispRel1a,lqgDispRel1b,lqgDispRel2} support of the possibility of a Planck-scale modified energy-momentum dispersion relation. In particular, two ways for the Planck-scale modification of the energy-momentum dispersion relation are considering in Refs. \cite{lqgDispRel1a,lqgDispRel1b,lqgDispRel2}: 
\begin{itemize}
\item The first one is $p(E)$ as an expansion with leading Planck-scale correction of order $L_p\, E^3$ \cite{Amelino2004b},
\begin{equation}
 \vec{p}^2 \simeq E^2 - m^2 + \alpha_1\, L_p\, E^3.
\label{disprelONE}
\end{equation}
\item The second one, in which the function $p(E)$ admits an expansion with leading Planck-scale correction of order $L_p^2 E^4$ \cite{Amelino2004b},
\begin{equation}
 \vec{p}^2 \simeq E^2 - m^2 + \alpha_2\, L_p^2\, E^4.
\label{disprelTWO}
\end{equation}
\end{itemize}

For a particle  of rest mass $M$, the position can be measured by a procedure involving a collision with a photon of energy $E$ and momentum $p$. Due to Heisenberg uncertainty principle, for determining the uncertainty in position $\delta x$ one should use a photon with momentum uncertainty $\delta p \ge 1/\delta x$.
On the other hand, on loop quantum gravity~\cite{landau}, $\delta p \ge 1/\delta x$. This can be converted into $\delta E \ge 1/\delta x$. By using the special relativistic dispersion relation, $\delta E \ge 1/\delta x$ and $M \ge \delta E $.

If  the loop quantum gravity indeed implies modified dispersion relation hosts at Planck scale. Eq. (\ref{disprelTWO}), this can be deduced from $\delta p_\gamma \ge 1/\delta x$. It turns to be necessary to have \cite{Amelino2004b},
\begin{equation}
M  \ge \frac{1}{\delta x} \left(1 - \alpha_2 \frac{3 L_p^2}{2 (\delta x)^2}\right).
\label{dis}
\end{equation}
It is obvious that these results are valid for a particle, which was at rest \cite{landau}. The generalization of these results to be applicable for the measurement of the position of a particle with energy $E$, is a straightforward task. In case of standard dispersion relation, one obtains that $E \ge 1/\delta x$, as required for a  linear dependence of entropy on area.
\begin{itemize}
\item For the dispersion relation, Eq. (\ref{disprelTWO}),
that 
\bea
E &\ge & \frac{1}{\delta x} \left(1- \alpha_2 \frac{3 L_p^2}{2
(\delta x)^2}\right),
\eea
which fulfils the requirements of derivation of a leading-order correction of log-area form.
\item For the dispersion relation, Eq. (\ref{disprelONE}), 
\bea
E  &\ge & \frac{1}{\delta x} \left( 1+ \alpha_1 \frac{L_p}{\delta x}\right),
\eea
\end{itemize}

In string theory, the proposed reversed Bekenstein argument leads,
\bea
\delta x &\ge & \frac{1}{\delta p} + \lambda_s^2  \delta.  \label{guphere}
\eea
This is a generalisation of the uncertainty principle. In Eq. (\ref{guphere}), the scale $\lambda_s$ is the effective string length, which apparently characterizes a length scale that might approach the Planck length.

\subsection{Modified dispersion relation in UHECR and TeV-GRB}
\label{sec:mdrUHE}

Furthermore, the energy-momentum uncertainty due to the quantum gravity origin the modification of the energy-momentum dispersion relation for such observations as ultra-high energy cosmic rays (UHECR) \cite{Lawrence:UHECRa,Lawrence:UHECRb,Lawrence:UHECRc} based on pion photo-production by inelastic collisions of cosmic-ray nucleons with the cosmic microwave background (CMB) $p + \gamma (CMB) \rightarrow p +\pi$, with  energy exceeding the Greisen-Zatsepin-Kuz’min (GZK) cutoff in order of $\times 10^{10}$ GeV. Another observation $20~$TeV $\gamma$--rays events \cite{Aharonian:TEV} for high energy photon propagating can suffer inelastic impacts with photons in the Infra-Red background resulting in production of an electron-positron pair $\gamma + \gamma (IR) \rightarrow e^+ + e^-$. Regarding very high energy, the uncertainty of energy--momentum reads as 
\bea
\delta E \gtrsim E \left(  \frac{E}{E_p}\right)^{a},\, \, \, \,  \delta p \gtrsim p \left(  \frac{p}{m_p c}\right)^{a}
\eea
where $a=1, 2/3$ corresponding to the fluctuations of the metric of such models $\delta g_{\mu \nu} \gtrsim l_p /l$ and $\delta g_{\mu \nu} \gtrsim (l_p /l)^{2/3}$ with respectively \cite{Y. J. Ng:2000}. For the energy exceeding the UHECR events and the $\gamma$-events in the scattering process, where incoming particle of momentum $p_1$ and energy $E_1$ collides with photon with energy $\omega$ tends to produce two energetic particle     with momenta $p_2$ and $p_3$ and energies $E_2$ and $E_3$, the energy-momentum dispersion relation reads as 
\bea
E_i =\sqrt{p_i ^2 + m_i ^2 },
\eea
where $i= 1,\,2,\,3$. The possible modification MDR already given \cite{Y. J. Ng:2000}, we consider the energy-momentum conservation
\bea
(E_i + \delta E_i )^2 = (p_i + \delta p_i )^2 + m_i ^2, 
\eea
for high energy $E_i \approx p_i$ becomes 
\bea
E_i \approx \frac{1}{2} p_i \left(2 + \frac{m_i ^2}{p_i ^2}+ \eta \frac{p_i ^2}{E_p ^a}\right),
\eea
where $\eta$ some parameter varying for different particle species and consider $\delta p_i - \delta E_i \approx \eta p_i ^{(1+a)}/ 2 E_p ^a$, the modified dispersion relation becomes \cite{Y. J. Ng:2000}
\bea
E_i \approx c^2 k^2  + \eta E^2 \left(\frac{E}{E_P}\right)^2,
\eea
where $p=c k$, $c$ speed of light, while the speed  of massless photon reads as\cite{Y. J. Ng:2000},
\bea
v = \frac{\partial E}{\partial k} \approx c\left[1 + \eta \frac{1}{2}(1+a) \left(\frac{E}{E_p}\right)^a \right]. 
\eea
The important result here, the photon speed is energy dependent. The energy-momentum dispersion relation for the individual particles participating in a collision with the UHECR and $\gamma$ TeV violated the Lorentz invariance energy-momentum conservation \cite{Y. J. Ng:2000}.  Although in the presence of the IR/UV mixing \cite{Matusis:UV2000,Douglas:UV2001,Amelino:UV2004} gives one sort of modified dispersion relations. 
\bea
m^2	\approx E^2 - \vec{p	}^2 + \frac{\alpha}{p^\mu \theta _{\mu \nu} \theta ^{\nu \sigma} p_\sigma} + \dots,
\eea
where $\alpha$ is parameter depend on various aspects of the field theory, for $\alpha \rightarrow 0$, the standard dispersion relation will be obtained. The photon dispersion relation in QG would be
\bea
p^2 c^2 + m^2 c^4= E \left\lbrace E+ \frac{E^2}{E_{QG}}\right\rbrace,
\eea
where the limit of $E_{QG}$ in order of the Planck energy,

\section{Space noncommutativity} 
\label{sec:nc}

In eliminating the point-like structure, the space noncommutativity (NC) was proposed \cite{NozariNC} as an alternative to GUP, and MDRs. Within noncommutative geometry, several attempts have been performed to find modification of Bekenstein-Hawking formalism of black hole thermodynamics \cite{NCBW1,NCBW2}. Accordingly, the evaporation of black hole ends at vanishing temperature and extremal remnant should have no curvature singularity.  The basic idea of  noncommutative constructions is that the commutator of two spacetime coordinates taken as operators, no longer vanishes \cite{NCBW2}. Space noncommutivity is conjectured to smear out the matter distributions on a scale associated with the so-called turn-on of noncommutativity. The smearing can be taken to be essentially Gaussian.

\subsection{Atomic structure}

The idea of noncommutativity has been triggered by results from string theory \cite{Douglas:UV2001,Wittena,Wittenb,Konechny:2002}. That the spacetime becomes noncommutative and gets an evidence of the necessity of spacetime quantization is to be originated in Ref \cite{Wittena,Wittenb,Snyder,8}. The noncommutative geometry in string theory within B-field and the string dynamic limits have been described by a minimally coupled gauge theories on a noncommutative space \cite{Wittena,Wittenb}. The NC spacetime structure in very special relativity satisfies the modified dispersion relation \cite{Gibbons:2008a,Gibbons:2008b}, point-particle Lagrangian \cite{Ghosh:2011} and also was introduced for charged particles in presence of an external electromagnetic interactions \cite{Ghosh:2012}. 

Noncommutativity of spacetime can be given as
\bea
\left[x^{\mu}, x^{\nu}\right]=i \theta ^{\mu \nu},
\label{NCSpace}
\eea 
where $\theta^{\mu \nu}$ is an anti-symmetric matrix determining the fundamental discretization and/or quantization of the phase space. The NC algebra of physical quantities is associated with a microscopic system based on de Broglie and Schroedinger wave equations. The latter possess all information about the structure of systems such as  hydrogen atom \cite{Alain:1989}. The description of such systems becomes simple when given in algebra of some observable quantities. The time evolution of these observable quantities are commutative and can be given as a series \cite{Alain:1989},
\bea
q(t) = \sum _{k} q_{1} \dots q_{k} e^{(2 \pi i \langle n, \nu\rangle t)}, 
\eea 
where $\nu _{i}$ is the fundamental frequency and $n_i$ is an integer. $\langle n, \nu\rangle = \sum _{i} n_i \nu _i$, and the time evaluation of the observed quantity can simply be obtained from the Heisenberg picture  (Poisson brackets $\lbrace a,b \rbrace=i \hbar \,[a,b]$), $\lbrace H,q \rbrace=dq(t)/dt$  \cite{Alain:1989}.
One defines $H$ as a particular physical quantity which plays an important role in defining the total  energy.
This is given by its coefficients $H_{(i,j)} = 0, \,\ni i \ne j, \, H_{(i,i)} = h \nu _{i}$, where $h$ is Planck's constant, i.e. a constant converting frequencies into energies. 

The results discussed so far show that a simplistic structure on a manifold such as phase space appears as deformation in the parameter $h$, while the algebra of the functions can be replaced by noncommutative algebra, for instance the gauge desecration of bosons in a noncommutative geometry  \cite{Dubois:1989a,Dubois:1989b}.

\subsection{Quantum field theory}

The framework of QFT differs from the one of QM. The QM Hamiltonian formulation allows to the representation in phase-space coordinates, while the one of the earlier formulates the Lagrangian. This difference does not allow the description of a system in both theories, straightforwardly, i.e. a method of treating the coordinates noncommutativity is needed \cite{Smailagic:2003c1,Smailagic:2003c2}. 

The existence of a minimal length determines the noncommutative and this in turn effects the noncommutativity by changing the nature of the coordinates, Eq. (\ref{NCSpace}), \cite{Smailagic:2003c1,Smailagic:2003c2}. The first test of a successful formulation of NC QFT tends to  UV/IR mixing and defies renormalization group expectations \cite{Smailagic:2003c1,Smailagic:2003c2}. In QFT framewrork, a mean value of a function of the  positions  $g(x_1, x_2)$ is assumed. When introducing the set of operators \cite{Smailagic:2003c1,Smailagic:2003c2},
\bea
\hat{Z} = \frac{1}{\sqrt{2}} (\hat{x_1}+i \hat{x_2}), \qquad \mathtt{and} \qquad
\hat{Z}^\dagger = \frac{1}{\sqrt{2}} (\hat{x_1}-i \hat{x_2}),
\eea
a quantum field on a noncommutative plane, $\hat{Z}$/$\hat{Z}^\dagger$ can be recognized as creation/annihilation operators, respectively, and satisfy the commutation relation $[\hat{Z}, \hat{Z}^\dagger]= \theta$. The corresponding eigenstates 
\bea
\hat{Z}|Z\rangle = z|Z\rangle, \qquad \mathtt{and} \qquad
\langle Z| \hat{Z}^\dagger  = \langle Z|\bar{z}.
\eea
The normalized states satisfy the coherent states of NC 
\bea 
|Z\rangle = \exp{\left(-\frac{z\bar{z}}{2 \theta}\right)}\, \exp{\left(-\frac{z}{\theta} \hat{Z}^\dagger \right)} |0 \rangle ,
\eea
where $| 0 \rangle$ is the vacuum state. The noncommutativity and plane wave, respectively, can be defined as  \cite{Smailagic:2003c1,Smailagic:2003c2},
\bea
g(z) &=& \int \frac{d^2p}{(2\pi)^2} f(p) \langle z|\exp{(i \,p_j \hat{x}^j)}|z\rangle, \\
\langle z|\exp{(i \,p_j \hat{x}^j)}|z\rangle &=& \langle z| \exp{(i\, p_+\, \hat{Z}^\dagger)} \exp{(p_-\hat{Z}|)} \exp{\left( \frac{p_- p_+}{2} [\hat{Z}^\dagger , Z] \right) }|z\rangle, 
\eea
where $p_{\pm}\equiv(p_1\pm ip_2)/\sqrt{2}$.
The plane wave of noncommutative generates a Gaussian distribution \cite{Smailagic:2003c1,Smailagic:2003c2},
\bea
g(z) &=& \frac{4\pi}{\theta} \exp{\left(-\frac{4}{\theta} z\bar{z}\right)},
\eea 
where the function $f(p)=const.$ expresses maximum momentum, which is related to the minimal length uncertainty and proportional to $\sqrt{\theta}$. The commutative space leads to $\theta \rightarrow 0$.

The solution can be estimated by the Dirac delta function. Another important result of Ref. \cite{Smailagic:2003c1,Smailagic:2003c2}  is the establish of the Feynman propagator in momentum space. In the same way, one can redefine the Feynman propagator of the path integral  for non-relativistic free particle in the  noncommutative plane \cite{Smailagic:2003c1,Smailagic:2003c2}
\bea
G_{\theta} (x-y; E)= \int \frac{d^2p}{(2 \pi)^2} \exp{\lbrace i\vec{p}(\vec{x}-\vec{y})\rbrace} G_{\theta}(E; \vec{p}^2),
\eea
where the Green function in the momentum space $G_{\theta}(E; \vec{p}^2) = (2\pi)^{-2} \exp{(-\theta \vec{p}^2 /2)}/[E+ \frac{\vec{p}^2}{2m}]$ \cite{Smailagic:2003c1,Smailagic:2003c2} has a large momentum cut-off of space $\theta$. Furthermore, the relativistic limit gives,
\bea
G_{\theta} (x-y; m^2) &=& N \int D_x\, D_p\, D_e\, \nn \\
&& \exp\left\{ i \int ^{x_2} _{x_1} p_{\mu} dx^{\mu}-\int _0 ^T \tau \left( e(\tau) (p^2 + m^2) + \frac{\theta\, \vec{p}^2}{2T}   \right) \right\},  
\eea
where $e(\tau)$ is a Lagrange multiplier, which includes the proper time $t$ so that $\int ^\infty _0 dt \delta \left[t- \int _0 ^T d \tau e(\tau)\right] = 1$ \cite{Smailagic:2003c1,Smailagic:2003c2}.
Finally, the relativistic form reads,
\bea
G_{\theta} (x-y; m^2) &=& N \int _0^\infty dt e^{-t m^2} \int \frac{d^3 p}{(2\pi)^3} e^ i p_\mu (x^\mu- y^\mu) \exp{\left[ - (t+\theta) \frac{\vec{p}^2}{2}\right]} \nn \\ 
&\equiv & \int _0 ^\infty \frac{d^3p}{(2\pi)^3} e^{i p_\mu (x^\mu- y^\mu)} G_{\theta} (p^2; m^2).
\eea
The corresponding Green function is $G_{\theta} (p^2; m^2) =(2\pi)^{-3}\, \exp{(- \theta \vec{p}^2 /2)}/[p^2 + m^2]$ \cite{Smailagic:2003c1,Smailagic:2003c2}.

\subsection{Noncommutativity space algebra}

The HUP is strongly related to the canonical commutation or the commutative phase-space structures. When HUP should be broken down due to GUP, the GUP being compatible to the string theory or the quantum gravity and excepted to presence of the minimal length scale or maximum momentum scale, the precision of the position coordinate tends to  such a large accumulation of momentum coordinate or energy density on the canonical uncertainty that the latter can appreciably alter the space-time metric. An operational form of the noncommutative (NC) phase-space structures shall be observed. The generic expressions were introduced in Ref. \cite{Kempf1994a,Kempf1994b,Subir2}
\bea
\left[x_{i},\, p_{j}\right] &=& i \hbar\, \left[  \delta_{i j}\, \left(1+\beta\, f_{1} \left(\textbf{p}^{2}\right) \right)+f_{2}\, \left(\textbf{p}^{2}\right)\; p_{i}\, p_{j}\right],\\
\left[x_{i},\, x_{j} \right] &=& i \hbar f_{i j}(p) \ne 0.
\eea
The presence of a minimum length  or a maximum momentum or both of them apparently leads to GUP originated in  the NC algebras. Accordingly, Kempf \cite{17} proposed the following algebraic relations: 
\bea
\left[x_{i}, p_{j}\right] &=& i\, \hbar\, \left[   \delta_{i j}\, \left( 1+\beta\, p^{2} \right)+ \beta^{'}\, p_{i}\, p_{j}\right],\\
\left[x_{i}, x_{j}\right] &=& i\, \hbar\, \left(\beta ^{'} - 2\, \beta \right) \left(x_{i}\, p_{j} - x_{j}\, p_{i} \right),\\
\left[p_{i}, p_{j}\right] &=& 0.
\eea
Other algebraic relations were introduced in Ref. \cite{16}
\bea
\left[x_{i}, p_{j} \right] &=& i\, \hbar\,  \delta_{i j}\, \left(1 + \beta\, p^{2}\right),\\
\left[x_{i}, x_{j} \right] &=& - 2\, i\, \hbar\, \beta\, \left(x_{i}\, p_{j} - x_{j}\, p_{i} \right),\\
\left[p_{i}, p_{j} \right] &=& 0.
\eea
Recent algebraic relations have been presented \cite{chang2q}
\bea
\left[x_{i}, p_{j} \right] &=& i\, \hbar\,  \left[   \delta_{i j}\, \left( 1+\beta\, p^{2} \right)+ \beta^{'} p_{i} p_{j}+O(\beta^{'2}, \beta^{2}) \right] ,\\
\left[x_{i}, x_{j} \right] &=& i\, \hbar\, \frac{\left(2\, \beta - \beta^{'} \right)+\left(2\, \beta + \beta^{'} \right)\,\beta\, p^{2}}{1+\beta\, p^{2}} \, \left( x_{i}\, p_{j} - x_{j}\, p_{i} \right),\\
\left[p_{i}, p_{j} \right] &=& 0.
\eea

The GUP approaches which are  consistent with the NC algebras offer the possibility for space discreteness and/or quantization. In other words, the physical states of space should be non-commute. Despite that the physical states can not be measured, simultaneously, the space discretization seems to be possible.

\section{Lorentz invariance violation}
\label{sec:liv}

The suggestion that the Lorentz invariance (LI) principle may represent an approximate symmetry of nature dates back to about four decades \cite{LI1a,LI1b}. When studying the Compton wavelength of the particle of interest, the Heisenberg uncertainty principle combined with finiteness of the speed of light $c$ leads to creation and annihilation processes \cite{garay}. The space-time foamy structure at small scales comes up with LI violation (LIV) as a likely consequence. A self-consistent framework for analysing possible LIV was suggested by Coleman and Glashow \cite{cg1,Glashow:2}. In gamma ray bursts (GRB), the energy dependent time offsets were investigated from standard cosmological model   in different energy bands \cite{jellis1a,jellis1b}. An energy dependent modification of the standard relativistic dispersion relation, section \ref{sec:mdr}, is seen as a manifestation for LIV. The redshift dependence of the time delays due to LIV has been found very weak. A comprehensive review on the main theoretical motivations and observational constraints on Planck scale suppressed LIV is given in Ref. \cite{revw} and the references therein. At energies approaching the Planck scale, various theoretical indications (such as quantum gravity scenarios and loop quantum gravity \cite{Ashtekar}) that LI breaks down because of the need to cut-off the UV divergences in QFT \cite{Rovelli}. Studying the Planck scale itself turns to be accessible in quantum optics \cite{nature2012}.

In General Relativity, the local Lorentz invariance principle representing rotations and boosts which are local symmetries of nature and the weak equivalence principle stating that gravity is flavor independent are two ingredients of the Einstein equivalence principle. Deep understanding of gravity at all scales is strongly connected with experimental prove for these principles at all scales. The local Lorentz invariance is obviously limited to the matter-gravity couplings \cite{Bialey}. Following Coleman-Glashow recipe, an effective field theory for general local LIV can be formulated as a Lagrange density (Einstein Hilbert term, cosmological constant, and series of operators of increasing mass dimension). The latter represents corrections to known physics at attainable scales \cite{Bialey,Kostelecky}

Due to LIV, the dispersion relation of a photon having distant origin, momentum $p$ and energy $E$ would be a subject of a tiny modification \cite{Vasileiou} 
\begin{eqnarray}
E^2 &\eqsim & p^2\, c^2 \left[1-\sum_{n=1}^{\infty} \eta\, \left(\frac{E}{\cal E}\right)^n\right],
\end{eqnarray}
where ${\cal E}$ is an energy scale (Planck energy $\sim\sqrt{(\hbar\, c^5)/G} \eqsim 1.22\, 10^{19}~$GeV), at which gravity is to be quantized and $\eta$ stands for positive or negative LIV. Apparently, the lowest order term is expected to dominate the series, especially at $E \ll {\cal E}$. The photon propagation speed reads,
\begin{eqnarray}
v(E) &=& \frac{\partial E}{\partial p} \eqsim c \left[1- \eta \frac{n+1}{2} \left(\frac{E}{\cal E}\right)^n\right], \label{eq:vE1}
\end{eqnarray}

When two particles (photons) with two different energies $E_h > E_l$ are emitted at the same time and from the same distance location, they arrive on Earth with a time delay $\Delta t(E)$. The speed of arrival can be calculated from Eq. (\ref{eq:vE1}). We make a further step and want o estimate the possible GUP correction, section \ref{sec:gup1}. The momentum of such a particle would be a subject of a tiny modification so that the comoving momenta reads \cite{Tawfik:2012hz}
\bea
p &=& p \left(1-\alpha\, p_0 + 2\, \alpha^2\, p_0^2\right), \\
p^2 &=& p^2 \left(1-2\, \alpha p_0 + 10\, \alpha^2\, p_0^2\right),
\eea
where $p_{0}$ is the momentum at low energy and $\alpha_0$ is dimensionless parameter of order one. In this co-moving frame, the dispersion relation  is given as
\bea
E^2 &=& p^2\, c^2 \left(1-2\,\alpha\, p_0\right) + M^2\, c^4.
\eea
When taking into consideration a linear dependence of $p$ on $\alpha$ and ignoring the higher orders, then the Hamiltonian $\textbf{H}=\left(p^2\, c^2 - 2\, \alpha\, p^3 \, c^2 + M^2\, c^4 \right)^{1/2}$. By implementing the relation between comoving and physical momenta \hbox{$p_{\nu}=p_{\nu_{0}}(t_0)/a(t)$}, where $a$ is the scale factor, the velocity is
\bea
v(t) &=& \frac{1}{a(t)} \frac{P_{\nu{_0}}^2\, c^2 -3 \alpha\, P_{\nu{_0}}^2 \, c^2}{\left(P_{\nu{_0}}^2\, c^2 -2\,\alpha\,P_0^3 + M_{\nu}^2\, c^4\right)^{1/2}} \\
 &=& \frac{c}{a(t)} \left[1-2\alpha p_0 - {\cal A} + 
\alpha p_0 \left(2 {\cal A} - {\cal B} + {\cal B} \, {\cal A}\right)\right]. \label{eq:vt}
\eea 
where ${\cal A}=M^2 c^2/(2 p^2)$ and ${\cal B}=M^2 c^4/(p^2 c^2 + M^2 c^4)$. In the relativistic limit, $p\gg M$, the fourth and fifth terms in Eq. (\ref{eq:vt}) simply cancel each other 
\bea
v(z) &=& c\,(1+z)\left[1-2\, \alpha\, (1+z)\, p_{\nu_0}  - \frac{M_{\nu}^2\, c^2}{2 (1+z)^2 p_{\nu_0}^2} +
\alpha\, \frac{ M_{\nu}^4 c^4}{2\, (1+z)^3\, p_{\nu_0}^3} 
\right], \label{eq:vz}
\eea
where $z$ denotes the redshift. In getting this expression, $p_0$ is treated as a comoving momentum. Then, the change in the relative velocities
\bea \label{eq:dvz1}
\Delta v(z) &=& \alpha\,c\, \left(- 2\, (1+z)^2\, p_{\nu_0} + \frac{ M_{\nu}^4 c^4}{2\, (1+z)^2\, p_{\nu_0}^3} \right) - \frac{M_{\nu}^2\, c^2}{2 (1+z) p_{\nu_0}^2}. \label{eq:vz2}
\eea

Recent Fermi-Large Area Telescope observations of four bright gamma-ray bursts (GRBs) reveal robust and stringent constraints on the dependence of $c$ on energy (vacuum dispersion relation), which is a form of LIV \cite{Vasileiou}. Measure the helicity dependence of the propagation velocity of photons originating in distant cosmological objects is one of the experimental tests of LIV. Strong upper limits on the total degree of dispersion were determined. A high degree of polarization was observed in the prompt emission \cite{Laurent}.  The existing constraint on LIV arising from the phenomenon of vacuum birefringence, section \ref{sec:bire}, could be improved by four order of magnitude.

\section{Planck-scale induced birefringence effect}
\label{sec:bire}

Some QG models assign remarkable properties to the space-time at very short distances; near the Planck length, empty space may behave as  crystal, singly or doubly refractive. Measuring the space refractivity and birefringence induced by gravity is one of the experimental tests for LIV and properties of the possible minimal length. Furthermore, the birefringence effect is once of the constraints related to time of flight. 

In quantum electrodynamics (QED), gamma and electron are the relevant particles to test LIV. Thus, the dispersion relation of photon, sections \ref{sec:mdr} and \ref{sec:liv}, can be utilized in deriving the gravity-induced birefringence. The Myers-Pospelov model introduces an effective field theory for the QED Lagrangian \cite{mp},
\begin{eqnarray}
{\cal L} &=& -\frac{1}{4} F_{\mu \nu} F^{\mu \nu} + \frac{1}{2 E_p} n^{\alpha} F_{\alpha \delta} n^{\sigma} \partial_{\sigma} \left(n_{\beta} \varepsilon^{\beta \delta \gamma \lambda} F_{\gamma \lambda} \right), \label{eq:mp1}
\end{eqnarray}
with mass-dimension five corrections, the second term which is quadratic, gauge invariant, and not reducible to lower-dimension operators, nor to a total derivative. The mass-dimension five term is Lorentz invariant, except for $n^{\alpha}$ which is an external four-vector characterizing the preferred frame and violates LI.

Assuming a pure-time vector $n^{\alpha}= (n_0, 0, 0, 0)$, then Eq. (\ref{eq:mp1}) can be rewritten as
\begin{eqnarray}
{\cal L} &=& -\frac{1}{4} F_{\mu \nu} F^{\mu \nu} + \frac{\xi}{2 E_p} \varepsilon^{j\, k\, l}  F_{0  j}  \partial_0\,  F_{k\, l}. \label{eq:mp2}
\end{eqnarray}
Then, under boost transformation, it violates LI but preserves the space isotropy \cite{mp2a,mp2b}. $\xi$ can be taken of order one.

When focusing on birefringence (dispersion relation) emerging from Myers-Pospelov Lagrangian with $E=\hbar\, \omega$ and $p=\hbar\, k$, the photon dispersion cab be given by \cite{Laurent}
\begin{eqnarray}
\omega^2 &=& k^2 \pm 2\, \xi \frac{k^3}{M_p},
\end{eqnarray}
where the sign is determined by the photon chirality (circular polarization). During the propagation of linearly polarized photon, the chirality leads to rotation of the polarization (vacuum birefringence). The averaged rotation angle along distance $d$ is
\begin{eqnarray}
\Delta \theta (p) &=& \frac{d}{2} \left[\omega_+(k)-\omega_-(k)\right] \approx \xi\, \frac{d}{2}\, \frac{k^2}{M_p}, \label{eq:thta1}
\end{eqnarray}
where $\omega_{\pm} = |k|(1\pm \xi k/M_p)$. 
The right-hand side of Eq. (\ref{eq:thta1}), $\Delta \theta (p)$, and $d$ can be determined from astrophysical observations. Accordingly, an upper  limit can determined, for instance for GRB041219A
\begin{eqnarray}
 \xi &<& \frac{2}{d}\, \frac{M_p\, \Delta \theta (k)}{k_2^2-k_1^2} \approx 1.11^{-14}
\end{eqnarray}
Comparing this result with the relevant regime, $\xi \sim 1$, explains the importance of additional symmetries.

Among others, the black hole entropy is affected by GUP \cite{Tawfik:2015kga}. Furthermore, the associated quantum effects in entropic gravity would modify the Newtonian gravitational law \cite{Ali:2013ma}. Despite, the latter is negligibly small, the coupling to electromagnetism should be taken into consideration

\section{Origin of minimal measurable quantities}
\label{sec:orgn}

The {\it chronon}, a hypothetical fundamental or indivisible interval of time taking the value of the ratio between the diameter of the electron and the velocity of light proposed by Robert Levi  \cite{Levi} in 1927, would be considered as the fist minimum measurable time interval ($\sim 10^{-24}~$s) proposed.  Within this time interval, Special Relativity (SR) and QM are conjectured to unify in framework of QFT. The impossibility to resolve arbitrarily small structures with an object of finite extension has been observed in string theory \cite{4,st34,st35,st36}. The string scattering in {\it super-Planckian regime} would leads to GUP. This apparently prevents a localization to better than the String scale. 

Modes with energies exceeding the Planck scale have to be taken into account in calculating the emission rate. This is because of the infinite blueshift of photons approaching a black hole horizon. These trans-Planckian problems were discussed in 1970s \cite{entr3}. In 1995, Unruh suggested \cite{unruh25} a modification in the {\it dispersion relation} to deal with this difficulty. Therefore,  the smallest possible wavelength is the one that takes care of the {\it trans-Planckian problem}. Starting from a generalization of the Poincare algebra to a Hopf algebra, a modification in the commutators of the space-time coordinates to solve the {\it trans-Planckian problem} has been proposed \cite{MR28}. 

The peculiar role of gravity to test physics at short distances has been observed \cite{mead22}. It was showed that the role of gravity should not mean increasing in the Heisenberg measurement uncertainty. Snyder believed that the cut-off in momentum space should be a {\it ''distasteful arbitrary procedure''} \cite{Snyder}. Therefore, instead of cut-off, a modification of the {\it canonical commutation relations} of both position and momentum operators has been proposed. Accordingly, noncommutative space-time or modification of the commutation relations increase the Heisenberg uncertainty such that a smallest possible resolution of structures can be introduced. A minimal length scale does not need to be in conflict with the Lorentz invariance principle. 

Utilizing fundamental limits governing mass and size of any physical system to register time dates back to nearly six decades, quantum clock \cite{wigner57,wigner58} in measuring {\it distances}. This was given as constraints on smallest accuracy and maximum running time as a function of mass and position uncertainties. While Heisenberg uncertainty principle requires that only one single simultaneous measurement of both energy and time can be accurate, Wigner second constraint is more severe. 

That the gravity might not be a fundamental force dates back to Bronstein \cite{Gorelik}, i.e. gravity does not allow an arbitrarily high concentration of mass in a small region of space-time (Schwarzshild singularity \cite{Gorelik}). For a test-particle, the minimal measurable distance, the gravitational radius $G\, ρ\,V/c^2$, should by no means be larger than its linear dimensions $V^{1/3}$ \cite{Bronstein}. Thus, an upper bound on density $\rho\, \lesssim c^{2}/G\, V^{2/3}$ can be determined and the possibilities for measurements become even more restricted than from the commutation relations \cite{cr25,cr27}. A quantum theory of gravitation is thought to generalize the uncertainty relations to Christoffel symbols. Due to impossibility of concentrating mass in a region smaller than its Schwarzschild radius, uncertainties in measuring average values of Christoffel symbols have been introduced \cite{mead20}.  

Heisenberg found that Fermi theory of $\beta-$decay \cite{frmbta1,frmbta2} is non-normalizable and accordingly refined a fundamental minimal length, cut-off \cite{Heisenberg}. Later one, he also proposed an idea that QM with a minimal length scale would be able to account for the discrete mass spectrum of the elementary particles \cite{Heisenberg2}, i.e. singularities in QFT became better understood \cite{Sabine}. Due to Lorentz invariance principle, discrete approaches to space and time remained unappealing \cite{Sabine}.

Based on QFT and to overcome singularities in fundamental theories, a fundamental length was necessary, i.e. regularization such as \textit{cut-off} was used. Since cut-off would not be independent of the frame of reference, problems with the Lorentz invariance principle would appear. The finding that the effect of regularization with respect to cut-off should be the same as that of a fundamentally discrete space-time dates back to the 1930's \cite{Heisenberg2,Pauli}. A fundamental finite length or a maximum frequency was not unknown in these years \cite{exmp5,exmp3,exmp6,exmp7}. Therefore, the fundamental length was thought to be in the realm of subatomic physics, $10^{-15}~$m.

In founding minimal length, the main milestones can be summarized as follows \cite{Tawfik:2014zca}. 
\begin{itemize} 
\item Singularities in fundamental theories (such as $\beta-$decay) lead to cut-off a minimal length scale in QM.
\item Distasteful arbitrary cut-off procedure leads to modification in the canonical commutation relations of both position and momentum operators.
\item Gravity at short distance and {\it ''gedanken''} (thought) experiments  lead to various scenarios suggested for minimal length scale are connected with some gravitational aspects. 
\item Trans-Planckian problem (black hole thermodynamic properties) leads to modification in the dispersion relations. 
\item QM (QFT) with a minimal length scale leads to modifications of the canonical commutation relations in order to accommodate a minimal length scale.
\item String theory leads to generalized uncertainty principle (GUP) based on string scattering in the {\it super-Planckian regime}.
\end{itemize}

\subsection{String theory}
\label{ssec:1}

In order to guarantee QG consistency at Planck scale, a GUP approach was proposed by Amati {\it et al.} \cite{guppapers}. Analysis of ultra high-energy scatterings of strings played an essential role. Some interesting effects are compared to the ones, which have been found in {\it usual} field theories, especially the ones originating from the soft short-distance behavior of string theory \cite{guppapers}. The hard processes are studied at a short distance as in high-energy fixed-angle scatterings.  The latter are apparently not able to test distances shorter than the characteristic string length $\lambda_{s}=(\hbar \alpha)^{1/2}$, where $\alpha$ is the string tension. 

Another scale is dynamically generated. The $d$-dimensional gravitational Schwarzschild radius $R(E) \sim (G_N E)^{1/(D-3)}$ is conjectured to approach the string length $\lambda_s$ \cite{guppapers}. This depends on whether $R(E)$ smaller or greater than $\lambda_s$. If $R(E)>\lambda_{s}$, then new contributions at distances of the order of $R(E)$ appear. This indicates a classical gravitational instability, which can be attributed to the black hole formation. If the opposite should be the case ($R(E)<\lambda_{s}$), then their contributions are irrelevant. Obviously, there are no black holes with a radius smaller than the string length. In light of  this, the analysis of short distances can go on. It has been shown that the larger momentum transfers do not always correspond to shorter distances. Precisely, the analysis of the angle distance relationship suggests the existence of a scattering angle $\theta_M$. When the scattering should take place at $\theta<\theta_M$, then the relation between the interaction distance and the momentum transfer is the classical one, i.e. follows the Heisenberg relation with $q\sim\hbar/b$, where $b$ is the impact parameter. But when $\theta\gg\theta_M$, then the classical picture is no longer valid. An important new regime where $\langle q \rangle \sim b$ would be constructed. This suggests a modification of the uncertainty relation at the Planck scale \cite{guppapers}
\be 
\Delta x \sim \frac{\hbar}{\Delta p} +Y\, \alpha\, \Delta p,
\ee
where $Y$ is a suitable constant. Consequently, the existence of a minimal observable length of the order of String size $\lambda_{s}$ is likely.

Tools of {\it ''gedanken''} string collisions at the Planck energy have been very useful \cite{4,guppapers}.  In addition to these, the renormalization group analysis has been applied to the string \cite{5}.

\subsection{Black hole physics} 
\label{ssec:2}

Several works have been devoted to perform the uncertainty relations and their measurability bounds in QG \cite{3}. Thought experiments have been proposed to measure the area of the apparent horizon of a black hole \cite{3}. Accordingly, a generalization of the uncertainty principle was deduced, which agrees well with the one stemming from the string theories \cite{3,guppapers,2}.  A main physical ingredient was the Hawking radiation \cite{entr3}. The black hole approach to GUP, which is a rather model independent approach, agrees, especially in its functional form, with the one obtained in framework of the string theory. 

The thought experiment proceeds by observing the photons scattered by the studied black hole. The main physical hypothesis of the experiment is that the black hole emits Hawking radiation. Detecting the Hawking radiation, it turns to be possible to grab a black hole  ''image'' \cite{entr3}. Besides, measuring the direction of the propagating photons that are emitted at different angles and tracing them back, one can - in principle - locates the position of the black hole center \cite{entr3}. In such a way, the radius $R_{h}$ of the apparent horizon will be measured. Apparently, this measurement has two sources of uncertainty \cite{3}.  
\begin{itemize}
\item The first one is based on the fact that a photon with wavelength $\lambda$ cannot carry information about a more detailed scale than $\lambda$ itself \cite{3}. As in the classical Heisenberg analysis, the resolving power of the microscope gives the minimum error $\Delta x^{(1)} \sim \lambda/\sin \theta$, 
where $\theta$ is the scattering angle. Then, the final momenta should have the uncertainty $\Delta p\sim h \sin(\theta)/\lambda$. During the emission process, the mass of the black hole varies from $M$ to $M-\Delta M$ \cite{3}, where $\Delta M=h/(c\, \lambda)$. The radius of the horizon changes, accordingly. The corresponding uncertainty is intrinsic to the measurement.

For example, the metric element of Reissner black hole \cite{Carroll} is given as 
\be 
ds^{2} = \left(1 -\frac{2\, M\, G}{r}+\frac{G\, Q^2}{r^{2}}\right) d t^{2} - \left(1 -\frac{2\, M\, G}{r}+\frac{G Q^2}{r^{2}}\right)^{-1}\, d r^{2} - r^{2} d \Omega^{2}.
\ee
Also, the apparent horizon is defined as the outer boundary of a region of closed trapped surfaces. In spherical topology and Boyer-Lindquist coordinates \cite{BL1967}, the apparent horizon is located at  $r = R_{h}$ 
\be 
R_{h} = G\, M\, \left[1+\left(1- \frac{Q^{2}}{G\, {M^{2}}}\right)^{1/2}\right].
\ee
The Boyer-Lindquist coordinates  are a generalization of the coordinates used for the metric of a Schwarzschild black hole. This can be used to express the metric of a Kerr black hole \cite{kerr1963}. Accordingly, the line element for a black hole with mass $M$, angular momentum $J$, and charge $Q$ reads
\bea
d s^2 &=& -\frac{\Delta}{\Sigma} \left(d t - K\, \sin^2(\theta)\, d \phi\right)^2 + \frac{\sin^2(\theta)}{\Sigma}\, \left(\left(R^2+K^2\right) d \phi - K\, d t\right)^2 \nn \\
&+& \frac{\Sigma}{\Delta}\, d R^2 + \Sigma\, d \theta^2, \hspace*{10mm}
\eea
where $\Delta = R^2 - 2 M R + K^2 + Q^2$, $\Sigma = R^2 + K^2 \cos^2(\theta)$ and $K=J/M$.
In Boyer-Lindquist coordinates, the Hamiltonian of a test particle is separable in Kerr space-time. From Hamilton-Jacobi theory, a fourth constant of the motion can be derived. This is known as Carter's constant \cite{CC1968}

\item The second source of uncertainty is the case, when $1 - 2 G/r+G Q/r^{2}$ vanishes. In 1D and for $M\,\gg\,\Delta M$ and $Q^2= G\, M^2$, the position uncertainty reads
\bea 
\Delta x ^{(2)} &=& G M \pm \sqrt{G^{2} (M+\Delta M)^{2}-G Q^{2}}, \\
\Delta x ^{(2)} &>& G \sqrt{2 M \Delta M} \geq \frac{2G}{c^2} \Delta M=\frac{2G}{c^3} \frac{h}{\lambda}.
\eea
By means of inequality $\lambda/\sin \theta \geq \lambda$, the uncertainty in $\Delta x^{(2)}$ and the quantity itself can be combined, linearly 
\bea
\Delta x & \gtrsim & \lambda +\kappa\, \frac{l_p^{2}}{\lambda} \label{eq:f}
\eea
where $\Delta x \gtrsim \frac{\hbar}{\Delta p} + c~G  \Delta p$ or $\Delta x  \gtrsim \frac{\hbar}{\Delta p} + \beta  \Delta p$, with $\kappa$ is a constant. The other numerical constant $\beta$ cannot be predicted by the model-independent arguments presented so far. It is natural to investigate whether the relation given in  Eq. (\ref{eq:f})  reproduces what was obtained considering only a very specific measurement. This principle would assure that the results should have a more general validity in QG.
\end{itemize}

In a {\it gedanken} experiment of a micro 4-dimensional black hole \cite{7}, another approach has been deduced. This approach is given as function of time and energy. When position with a precision $\Delta x$ is measured, the quantum fluctuations of the metric field around the measured position with energy amplitude can be expected as $\Delta E \sim c\,  \hbar/(2\, \Delta x)$.
The Schwarzschild radius associated with the energy fluctuation $\Delta E$ is given as $R_s = 2\,  G_N\,  \Delta E/c^4$.
The energy fluctuation $\Delta E$ would grow up and the corresponding the radius $R_s$ would become larger and larger, until it reaches the same size as $\Delta x$. As it is well known, the critical length is the Planck length, $R_s = \Delta x \equiv l_p$, where $l_p ^{2}=G_N  \hbar/c^3$ and the associated energy is the Planck energy $\epsilon_p=\hbar\, c/(2\, l_p) =  \sqrt{\hbar\, c^5/G_N}/2$.

When the discussion is limited to the Planck energy, the Schwarzschild radius $R_s$ is considerably enlarged. The situation can be summarized by the inequalities
$\Delta x \gtrsim  \frac{c  \hbar}{2 \Delta E}  \Longrightarrow   \Delta E \ll \epsilon_p$ or $\Delta x \gtrsim 2  G_N  \Delta E/c^4  \Longrightarrow   \Delta E \sim \epsilon_p$.
If these two inequalities are combined linearly, then
\bea
\Delta x &\gtrsim&  \frac{c  \hbar}{2 \Delta E} + \frac{2  G_N  \Delta E}{c^4}. 
\eea
This is a generalization of the uncertainty principle to the cases in which gravity gets very important, i.e. to energies of order of $\epsilon_p$. We have discussed this in connection with the various colliders and the indirect relation between energy and wavelength. We noticed that this relation might be violated at very high energy due to the dominant role of gravity at this energy scale. It is obvious that the minimum value of $\Delta x$ is reached for $\Delta E_{max} \sim \epsilon_P$, 
$\Delta x_{min} = 2\, l_p$.

\subsection{Snyder form}
\label{ssec:3} 

A relationship between a dual structure and the associated product rules fulfilling certain compatibility conditions is introduced by the Hopf algebra \cite{MR28}. An additional structure was found in this geometric approach. The curvature of momentum-space is expressed in terms of coproducts and antipodes of the Hopf algebra \cite{MR28}. In light of this, a theory for quantized space-time was proposed \cite{8,9}. In resolving the infinities problem in early days of QFT different possibilities are investigated.  A de-Sitter space with real coordinates $(\eta_{0},\eta_{1},\eta_{2},\eta_{3},\eta_{4})$ was taken into account. By choosing different parametrizations of the hypersurface than the ones proposed in Ref. \cite{MR28}, one can also use different coordinates in the momentum-space. One such parametrizations, coordinates $\pi_{\nu}$ are related to Snyder basis \cite{MR28}:
\bea
\eta_{0} &=& - m_{p}\, \sinh\left(\frac{\pi_{0}}{m_{p}}\right) - \frac{\vec{\pi}^{2}}{2\, m_{p}} \exp\left(\frac{\pi_{0}}{m_{p}}\right),\\
\eta_{i} &=& - \pi_{i}\, \exp\left(\frac{\pi_{0}}{m_{p}}\right),\\
\eta_{4} &=& - m_{p}\, \cosh\left(\frac{\pi_{0}}{m_{p}}\right) - \frac{\vec{\pi}^{2}}{2\, m_{p}} \exp\left(\frac{\pi_{0}}{m_{p}}\right),
\eea
where on the hypersurface $\eta_{4}$ is not constant and $\pi_{\nu}$ is the bicrossproduct basis of the Hopf algebra \cite{MR28}.

The position $X$ and time $T$ operators, which act on functions of variables $(\eta _{0},\eta _{1},\eta_{2},\eta _{3},\eta _{4})$, respectively,  are defined as \cite{8,9} 
\bea
X_{i} &=& i\, a\, \left(\eta_{4}\, \frac{\partial}{\partial\, \eta_{i}} - \eta_{i}\, \frac{\partial}{\partial\, \eta_{4}}\right), \\ 
T &=& \frac{i\, a}{c}\left(\eta_{4}\, \frac{\partial}{\partial\, \eta_{i}} + \eta_{i}\, \frac{\partial}{\partial\, \eta_{4}}\right).
\eea 
where $i=1,2,3$ and $a$ is a natural unit of length. Also, the energy  $P_{i} = (\hbar/a)\, \eta_{i}/\eta_{4}$, and momentum operators
$P_{T} = (\hbar/a)\, \eta_{0}/\eta_{4}$  \cite{8,9}.
Thus, the commutators between positions and momenta read
\bea
\left[X_i,\, P_j\right] &=& i\, \hbar\, \left[1+\left(\frac{a}{\hbar}\right)^{2}\, P^{2}\right],
\eea
where $P^{2}=\sum_{j}^3 P_j\, P_j$.

\subsection{Doubly Special Relativity}
\label{ssec:5}

Doubly relativistic theories are group of transformations with two Lorentzian invariants \cite{12}, the constant speed of light and an invariant energy scale.  By parametrization with respect to an invariant length $l$, a nonlinear realization of Lorentz transformations ($E$, $p$) was proposed \cite{13}. Thus, the auxiliary-linearly transforming variables $\epsilon$, and  $\pi$, respectively, read
\bea
\epsilon &=& E\, f\left(l\, E, l^{2}\, p^2 \right), \\
\pi_{i} &=& P_i\, g\left(l\, E, l^{2}\, p^2 \right).
\eea
With rotations realized as linearly depending on the dimensional scale \cite{12}, the two functions $f$ and $g$ parametrize nonlinear realization of Lorentz transformations. Corresponding to the choice of $f$ and $g$ \cite{Amelinoa,Amelinob,garay1,Scardigli}, Lorentz transformations of energy-momentum of a particle in different inertial frames should differ from the transformations, which recover a nonlinear realization of the Lorentz transformation, when $l\, E \ll 1$ and $l^{2}\, p^{2}\ll 1$
\bea
f &=& \frac{1}{2} \left[\left(1+ l^2\, p^{2} \right) \frac{e^{l\, E}}{l\, E} -\frac{e^{-l\, E}}{l\, E} \right], \\
g &=& e^{l\, E}.
\eea
For a particle of mass $m$, the energy and momentum are related to  each other  $\left(1- l^2\, p^{2} \right) e^{l\, E}+e^{-l\, E}=e^{l\, m}+e^{-l\, m}$ \cite{Amelinoa,Amelinob,garay1,Scardigli}. Accordingly, $\exp(l\, E) = (\cosh(l\, m)+\sqrt{\cosh^{2}(l\, m)-\left(1-l^{2} p^{2}\right)})/\left(1-l^{2} p^{2}\right)$.
Furthermore, the upper bound on the momentum reads $p_{max}^{2}<1/l^{2}$. This suggests the existence of a minimal measurable length restricting the momentum to take any arbitrary value.  At the Planck scale, this leads to a maximal momentum due to the fundamental structure of space-time \cite{12}.

Following commutation relation given in Ref. \cite{12}
\bea
\left[X_i\,, P_j \right] &=& i\, \hbar\, \left[ e^{-l\, E} \delta_{i\, j} + \frac{l^2\,  p_{i} p_{j}}{\cosh {(l\, m)}} \right],
\eea
it is obvious that when the mass $m$ becomes much larger than the inverse of the length scale $l$, a classical phase-space is approached. This result obviously relates the transition from quantum to classical behavior with a corresponding modification in QM. The latter is induced by a modification of the relativity principle \cite{12}. 

If we consider massless particle, then $\exp(l\, E) =1/1- l\, |\textbf{p}|$ and the commutation relation should be modified \cite{12} 
\bea
\left[ X_i,\, P_j \right] &=& i\, \hbar\, \left[(1 - l\, |\textbf{p}|)\, \delta_{i j}+ l^2\,  p_{i}\, p_{j} \right]. \label{uuu}
\eea
When the momentum approaches its maximum value, a non-trivial limit for the canonical commutation relation shall be reached \cite{12}.

\section{Approaches for generalized (gravitational) uncertainty principle}
 \label{sec:gup1}
 
Based on the various GUP approaches \cite{Tawfik:2014zca}, the existence of a minimal length suggests that the space in the Hilbert space representation \cite{16} describes a noncommutative geometry, which can also arise as a momentum over curved spaces \cite{Kempf1994a}. From various {\it gedanken} experiments designed to measure the area of the apparent horizon of a black hole in QG \cite{20}, the uncertainty relation was preformed \cite{3}. The modified Heisenberg algebra introduces a relation between QG and Poincare algebra \cite{20}. In an $n$-dimensional space and under the effects of GUP, it is found that even the gravitational constant $G$ \cite{Extra} and the Newtonian law of gravity \cite{7} are subject of modifications. The interpretation of QM through a quantization in $8$-dimensional manifold implies the existence of an upper limit in the accelerated particles \cite{21}. Nevertheless, the quadratic and linear GUP approaches \cite{3,16,12} assume that the momenta approach the maximum value at very high energy (Planck scale)  \cite{12}. 

Another GUP approach fits well with the string theory and the black hole physics (with quadratic term of momenta) and with doubly special relativity (DSR) (with linear term of momenta)  \cite{advplb}. This approach predicts a minimal measurable length and a maximum measurable momentum, simultaneously and suggests that the space should be quantized and/or discritized. But, it has severe difficulties discussed in Refs. \cite{pedrama,pedramb}. Therefore, 
\begin{itemize}
\item a new GUP approach is conjectured to absolve an extensive comparison with Kempf, Mangano and Mann (KMM) \cite{16} and
\item another GUP approach was introduced to characterize a minimal length uncertainty and a maximal momentum, simultaneously, \cite{pedrama,pedramb}.
\end{itemize}
The latter has been performed in Hilbert space \cite{Nouicer}. Here, a novel idea of minimal length modelled in terms of the quantized space-time was implemented. Thus, this new approach agrees well with quantum field theory (QFT) and Heisenberg algebra, especially in context of non-commutative coherent states representation. The resulting GUP approach can be studied at ultra-violet (UV) finiteness of Feynman propagator \cite{Nouicer}.

The Quantum Gravity (QG) describes the quantum behaviour of gravitational field and unifies  the  Quantum Mechanics (QM) with the General Relativity (GR). As we discussed in previous sections, there are different approaches such as string theory, black hole physics and double special relativity, in which likely the Heisenberg uncertainty principle (HUP) is conjectured to be violated. Accordingly, various quantum mechanical systems would be subjects of modification.

The consistent unification of the classical description of GR with QM still an open problem. One attempt assumes that the two theories can be used as a guiding principle to the search of a fundamental theory of QG. Another one gives several arguments ranging from theoretical analysis in string theory to more sophisticated or even {\it ''gedanken''} experiments in order to measure the minimal length. Accordingly, a new contribution to the quantum uncertainty with a gravitational origin leading to a length scale as a Planck length  in the determination of space-time coordinates can be concluded. 

Various observations point to the applicability of the different GUP approaches towards interpreting the influences of the minimal length on the properties of a wide range of physical systems, especially at quantum level  \cite{3,7,Scardigli}. The effects of linear GUP approach have been studied on 
\begin{itemize}
\item recent cosmic inflation observations \cite{Tawfik:2014dza},
\item Newtonian law of gravity \cite{Ali:2013ma}, 
\item Inflationary parameters and thermodynamics of the early Universe \cite{Tawfik:2012he}, 
\item Physics of compact stars \cite{Ali:2013ii}, 
\item Lorentz invariance violation \cite{Tawfik:2012hz} and 
\item Measurable maximum energy and minimum time interval \cite{DahabTaw}. 
\end{itemize}
Regardless some applicability constraints, the effects of QG on the quark-gluon plasma (QGP) are also introduced, as well  \cite{Elmashad:2012mq}. It was found that the GUP can potentially explain the small observed violations of the weak equivalence principle in neutron interferometry experiments \cite{expa,expb,expc}, and also predicts a modified invariant phase space which is relevant to the Lorentz transformation. It was suggested \cite{nature2012} that GUP can be measured directly in quantum optics laboratories \cite{Das1,afa2}. Furthermore, deformed commutation relations would cause new difficulties in quantum as well as in classical mechanics. We give a list of some of these problems as follows. 
\begin{itemize}
\item 1-dimensional harmonic oscillator with minimal uncertainty in position \cite{16} and minimal uncertainty in position and momentum \cite{sssa,sssb} and $d$-dimensional harmonic oscillator with position minimal uncertainty  \cite{ddd,www},
\item problem of $3$-dimensional Dirac oscillator \cite{w}  and the solution of ($1+1$)-dimensional Dirac oscillator within  Lorentz covariant algebra \cite{ww}, 
\item $1$- and $3$-dimensional Coulomb problem within deformed Heisenberg algebra in perturbation theory  \cite{22,23,24,25,26},
\item scattering problem in deformed space with minimal length \cite{29},
\item ultra-cold neutrons in gravitational field with minimal length \cite{30,31,32},
\item influence of minimal length on Lamb shift, Landau levels, and tunnelling current in scanning tunnelling microscope \cite{afa2,Das}
\item Casimir effect in a space with minimal length \cite{35},
\item effect of non-commutativity and the existence of a minimal length on the phase space of cosmological model \cite{36},
\item various physical consequences of non-commutative Snyder space-time geometry \cite{37}, and 
\item classical mechanics in a space with deformed Poisson brackets  \cite{chang2q,39,40}.
\end{itemize}

On one hand, these approaches provide essential predictions. DSR suggests a possibility to relate the transition from the quantum behavior at the microscopic level to the classical behavior at the macroscopic level with the modification of QM induced by a modification of the relativity principles. Thus, the laboratory tests should be able to judge about these theories. On the other hand, the predictions remain uncertain due to the limitations of the current technologies. Nevertheless, the minimal length has been observed in condensed matter and atomic physics experiments, such as Lamb shift \cite{Das,Das1}, Landau levels \cite{Das,Das1}, and the Scanning Tunnelling Microscope (STM) \cite{Das1}.

As discussed, it seems that HUP likely breaks down at energies close to the Planck scale. Taking into account the gravitational effects, an emergence of a minimal measurable distance seems to be inevitable. More generally, the generalized (gravitational) uncertainty principle (GUP)  can be expressed as \cite{16}
\be 
\Delta x  \Delta p \geq \frac{\hbar}{2} \left(1+\alpha (\Delta x)^2 + \beta (\Delta p)^{2} +\zeta \right),
\ee
where both $\beta$ and $\zeta$ are positive and independent variables. The uncertainties in position $ \Delta x $ and momentum $\Delta p$ may depend on the expectation values of the operators $\textbf{x}$ and $\textbf{p}$, respectively; $\zeta= \alpha \langle x \rangle^2 +\beta \langle p \rangle^2$. 

According to HUP, the position minimal uncertainty $\Delta x_0 \ne 0$ is finite but  $\Delta x_{min} \propto \Delta p_{max}$ is proportional \cite{16}. Therefore, $\left[\textbf{x},\textbf{p}\right] = i\, \hbar\, \left( 1+\alpha\, \textbf{x}^{2}+\beta\, \textbf{p}^{2}\right)$ describes the resulting commutation relation.  
In QM, both $\textbf{x}$ and $\textbf{p}$ can be represented as operators acting on position-  and momentum-space wavefunctions, $\phi(x)=\langle x| \phi (x) \rangle$ and $\phi(p)=\langle p| \phi (p) \rangle$, respectively, where $|x\rangle$ and $|p\rangle$ are the position and momentum eigenstates, respectively. Both operators $\textbf{x}$ and $\textbf{p}$ are essentially self-adjoint. Their eigenstates can be approximated to an arbitrary precision by sequences of the physical states $|\phi _n \rangle$ of the increasing localization in position- $\lim_{n\rightarrow\infty} \Delta x_{|\phi _n \rangle}=0$ or momentum-space $\lim_{n\rightarrow\infty} \Delta p_{|\phi _n \rangle}=0$.

As pointed out in Refs. \cite{14,15}, with the inclusion of minimal uncertainties $\Delta x_0 >0$ and/or  $\Delta p_0 >0$, this situation changes, drastically. For example, a non-vanishing minimal uncertainty in position is given as $(\Delta x)_{|\phi _n \rangle}^{2} = \langle \phi | \left(\textbf{x}-\langle \phi | x |\phi \rangle \right)^{2}| \phi \rangle \geq \Delta x_{0}, \longrightarrow |\phi  \rangle$,  implying that no physical state would exist with such a position eigenstate \cite{16}. This is because an eigenstate would  of course have vanishing position uncertainties. It is apparent that a minimal position uncertainty means that the position operator is no longer essentially self-adjoint but symmetric. The preservation of symmetry assures that all expectation values should be real. When self-adjointness is abandoned, the introduction of minimal uncertainties is likely \cite{16}. 

Because of the absence of position eigenstates $|x\rangle$ in representation of the Heisenberg algebra, the Heisenberg algebra no longer finds Hilbert space representation on the position wavefunctions $\langle x| \phi (x) \rangle$ \cite{16}. In light of this, the discussion should be restricted to $\Delta x_{0} \ne 0$ and therefore $\alpha=0$, where there is no minimal momentum uncertainty. Similarly, a minimal momentum uncertainty is conjectured to abandon the momentum space wavefunctions \cite{16}. This allows to work with the convenient representation of the commutation relations on the momentum space wavefunctions
\bea
\Delta x  \Delta p &\geq& \frac{\hbar}{2} \left(1+ \beta (\Delta p)^{2} +\zeta \right),
\eea
where the constant $\zeta$ is positive and related to the expectation value of the momentum, $\zeta = \beta \langle p \rangle^{2}$.

\section{Minimal length uncertainty} 
\label{sec:min1}
 
Due to HUP, it exists no restriction on the measurement precision for the particle's position, $\Delta x$. This minimal position uncertainty can be made arbitrarily small even down to zero \cite{Scardigli}. The theoretical argumentation to avoid such a limit is reviewed in earlier sections. It is obvious that going down to such a limit is not essentially the case of the framework of GUP, because of the existence of a minimal length uncertainty, which obviously modifies the Hamiltonian of the physical system leading to modifications, especially at the Planck scale, in the energy spectrum of the quantum system, which in turn predict small corrections in the measurable quantities. As discussed in section \ref{sec:gup1}, this has been observed in condensed matter and atomic physics experiments, such as Lamb shift \cite{Das1,Das}, Landau levels \cite{Das1,Das}, and the Scanning Tunnelling Microscope (STM) \cite{Das1}. Thus, a hope arises that the quantum gravity effects may be observable in the laboratory.

We review two GUP approaches suggesting the existence of minimal length uncertainty. We summarize the mean features to each of them in Tab. \ref{tab:1}. In section \ref{sec:mm1}, we show the proposal of the minimal length uncertainty with momentum modification  \cite{Das1,Das}. In section \ref{sec:hs1}, we study the main features in Hilbert space representation of QM for the minimal length uncertainty \cite{16}.

\subsection{Momentum modification} 
\label{sec:mm1}

Via Jacobi identity, the GUP approach modifies the Heisenberg algebra as follows.
\bea  \label{veg}
\left[x_{i},\, p_{j}\right]=i \, \hbar\, \left(\delta _{i j}(1+\beta\, p^{2}) + 2\, \beta\, p_{i}\, p_{j} \right),
\eea
This ensures \cite{Das,Das1} that $\left[x_{i},\, x_{j}\right]=\left[p_{i},\, p_{j}\right]=0$. Thus, both position and momentum operators read  
\bea 
X_{i} &=& x_{0 i}, \label{dass}\\  
P_{j} &=& p_{0 j}\, (1+\beta\, p_{0}^{2}). \label{dasss}
\eea
It is obvious that $p_{0}^{2} = \sum_{j}^3 p_{0 j} p_{0 j}$ satisfies the canonical commutation relations $\left[x_{0 i},\, p_{0 j}\right] = i\, \hbar\, \delta_{i j}$ and $p_{0 j}$ is defined as the momentum at low-energy scale; $p_{0 j}=-i\, \hbar(\de/\de\, x_{0 j})$, while $P_{j}$ is considered as the momentum at high-energy scale.

As discussed earlier, the introduction of a minimal length leads to modification in the canonical commutation relations, while the position space at the Planck scale must differ from the position in the canonical system, because the absence of zero-state in the position eigenstates. Thus, it is useful to modify the position space rather to allow for modification in momentum space. The latter leads to non-commutation of space $\left[x_{i},\, x_{j}\right] \ne 0$.

From the assumptions given in Eqs. (\ref{dass}) and (\ref{dasss}), it is impossible to utilize Hilbert representation for the position space, since no zero physical state exists. With  the definition of the modified  momentum at the highest energy scales, Eq.  (\ref{dasss}), the non-commutative values of the momentum states $\left[p_{i},\, p_{j}\right] \ne 0$. We conclude that {\bf this approach fails to be represented in the Hilbert space}.

\subsection{Hilbert space representation} 
\label{sec:hs1}

We discuss a generalized framework to implement the appearance of a non-zero minimal uncertainty in the position. The discussion can be confined to exploring the applications of such a minimal uncertainty in the context of non-relativistic QM. Various features of the Hilbert space representation of QM, especially at the Planck scale, were introduced \cite{16}.
\bea 
\Delta x\, \Delta p &\geq & \frac{\hbar}{2}+ \beta_0\, l_p^{2}\, \frac{(\Delta p)^{2}}{\hbar ^{2}}. 
\eea
The second term, $\beta_0\, l_p ^{2}\, (\Delta p)^{2}/\hbar ^{2}$, finds its origin in nature of the spacetime at the Planck energy $\epsilon_p $  (of $10^{39}$ GeV) \cite{16,Scardigli}. The simplest GUP approach implies the appearance of a non-zero minimal uncertainty $\Delta x_0$  
\bea 
\Delta x\, \Delta p \geq \frac{\hbar}{2} \left(1+ \beta\, (\Delta p)^{2} \right),  \label{eg}
\eea
where $\beta = \beta_0/(M_p\, c^2) =\beta_0\, l_p ^2/\hbar ^2$ is the GUP parameter.

As a non-trivial assumption,  the minimal observable length is conjectured to have a minimal but non-zero uncertainty. Therefore,  the Hilbert space representation on position space wavefunctions of ordinary QM \cite{16} is no longer possible, as no physical system with a vanishing position eigenstate $|x \rangle$ is allowed \cite{16}. In light of this, a new Hilbert space representation which should be compatible with the commutation relation in GUP, Eq. (\ref{eg}), must be constructed. This means working with the convenient representation of the commutation relations on momentum space wavefunctions \cite{16}. Accordingly, the Heisenberg algebra of GUP is given as \cite{3,16,Scardigli,14,15,gupps2,Inflation2q,kmpf32}

\bea 
\left[x,\, p\right] &=& i\, \hbar\, \left(1+\beta\, p^{2}\right). \label{KMM1}
\eea
The Heisenberg algebra can be represented in the momentum space wavefunctions $\phi(p)=\langle p| \phi (p) \rangle $ and $\partial_{p}=i \hbar (\partial/\partial x)$
\bea 
\textbf{P}\, \cdot \phi(p) &=& p\, \phi(p), \label{eq:app} \\ 
\textbf{X}\, \cdot \phi(p) &=& i\, \hbar\, \left(1+\beta\, p^{2}\right) \partial_{p} \phi(p), \label{eq:app1}
\eea
where $\textbf{X}$ and $\textbf{P}$ are symmetric operators on the dense domain $S_{\infty}$ with respect to the scalar product $\langle \phi | \psi \rangle = \int_{-\infty}^{\infty} \frac{dp}{1+\beta p^2} \phi^{*} (p) \psi (p)$, the identity operator $\int_{-\infty}^{\infty} \frac{dp}{1+\beta\, p^2} | p \rangle \langle p | = 1$ and the scalar product of the momentum eigenstates changes to $
\langle p | p^{'} \rangle = \left(1+ \beta\,  p^{2} \right) \delta \left( p - p^{'} \right)$.
While the momentum operator  essentially still self-adjoint, the functional analysis of the position operator as expected from the appearance of the minimal uncertainty in positions should be changed. For $ (\Delta\, p)^{2} = \langle p^{2}\rangle -\langle p\rangle^{2}$ \cite{16}
\bea 
\Delta x\, \Delta p &\geq & \frac{\hbar}{2}\left(1+\beta\, (\Delta p)^{2} +\beta\, \langle p \rangle ^{2} \right).
\eea
This relation can be rewritten as a second-order equation for $\Delta p$. Then, the solutions for $\Delta p$ are  \cite{16}
\bea
\Delta p &=& \left(\frac{\Delta x}{\hbar\, \beta}\right) \pm \sqrt{\left(\frac{\Delta\, x}{\hbar\,  \beta}\right)^{2}-\frac{1}{\beta}-\langle p \rangle^{2}}.
\eea
A minimum position uncertainty $\Delta x_{min} (\langle p \rangle) = \hbar\, \sqrt{\beta} \sqrt{1+\beta\, \langle p \rangle ^{2}}$.
Therefore, the absolutely smallest uncertainty in position, where $\langle p \rangle =0$, $\Delta x_{0}=\hbar\, \sqrt{\beta}$.
There is a non-vanishing minimal momentum uncertainty.

For Hilbert space representations, one has to resort a generalized Bargmann-Fock representation \cite{fock,fock1} instead of working on position space. Here, the situation with non-zero minimal position uncertainties should be specified. At $n$-dimensions, the generalised Heisenberg algebra, Eq. (\ref{eg}), reads \cite{3,16,Scardigli,14,15,gupps2,Inflation2q,kmpf32}
\be 
\left[x_{i},\, p_{j}\right]= i\, \hbar\,  \left(1+\beta\, \vec{p}^{2} \right) \label{egu1}
\ee
which requires that 
\be 
\left[p_{i},\, p_{j} \right] = 0,  \label{egu2}
\ee
in order to allow a generalization of the momentum space representation \cite{16}
\bea
\textbf{P}_{i} \cdot \phi(p) &=& p_{i}\, \phi(p),\\
\textbf{X}_{i} \cdot \phi(p) &=& i\, \hbar\, \left(1+\beta \vec{\textbf{p}}^{2} \right) \partial_{p_{i}} \phi(p),
\eea
and $\partial_{p_{i}}=i\, \hbar\, (\partial/\partial p_{i})$. It turns to be obvious that
\bea
\left[{\bf X}_{i},\, {\bf X}_{j} \right] &=& 2\, i\, \hbar\, \beta\, \left({\bf P}_{i}\, {\bf X}_{j} - {\bf P}_{j}\, {\bf X}_{i} \right), \label{egu3} 
\eea
leads to a non-commutative geometric generalization of the position space.

Furthermore, the commutation relations, Eqs. (\ref{egu1}), (\ref{egu2}) and (\ref{egu3}) do not violate the rotational symmetry \cite{16}. In fact, the rotation generators can be expressed in terms of position and momentum operators  \cite{16}
${\bf L}_{i j} = ({\bf X}_{i}\, {\bf P}_{j} -  {\bf X}_{j}\, {\bf P}_{i})/(1+ \beta\, \vec{\textbf{p}}^{2})$, where their representation in momentum wavefunctions ${\bf L}_{i j}\, \psi(p) = -\, \, i\, \hbar\, \left( p_{i}\, \partial_{p_{j}} - p_{j}\, \partial_{p_{i}}\right)\, \psi(p)$
are essentially the same as encountered in ordinary QM. However, the main change now appears in the relation
\be
\left[x_{i},\, x_{j} \right] = -\, 2\, i\, \hbar\, \beta\, \left( 1+ \beta\, \vec{\textbf{p}}^{2} \right)\; L_{i\, j}.
\ee
Once again, this relation reflects the noncommutative nature of the spacetime manifold at the Planck scale.

\subsubsection{Eigenstates of position operator in momentum space}

The position operators generating momentum-space eigenstates are given as \cite{16}
\bea 
\textbf{X} \, \phi_{\lambda} (p) &=& \lambda\, \phi_{\lambda}(p),\\
i\, \hbar\, \left(1+\beta\, p^{2}\right)\, \partial_{p}\, \phi_{\lambda}(p) &=& \lambda\, \phi_{\lambda}(p).
\eea
This differential equation can be solved to obtain formal position eigenvectors $\phi_{\lambda} (p) = C\, \exp(-i \lambda/(\hbar\, \sqrt{\beta})\, \tan^{-1}\, \sqrt{\beta}\, p)$ \cite{16}.  By applying the normalization condition, the formal position eigenvectors in momentum-space can be found $
\phi_{\lambda} (p) = \sqrt{\sqrt{\beta}/\pi} \exp(- i\, \lambda/(\hbar\, \sqrt{\beta}) \tan^{-1} \sqrt{\beta}\, p)$  \cite{16}. 
This is the generalized momentum-space eigenstate of the position operator in the presence of both a minimal length and a maximal momentum. To this end, we calculate the scalar product of the momentum space eigenstate of the position operator $|\phi_{\lambda}(p)\rangle$  \cite{16},
\bea
\langle \phi_{\lambda^{'}} |\phi_{\lambda} \rangle &=& \sqrt{\beta}/\pi\, \int_{-\infty}^{\infty} \frac{dp}{1+\beta p^2}\, \frac{\exp(-i (\lambda-\lambda^{'})}{\hbar \sqrt{\beta}) \tan^{-1} \sqrt{\beta} p} \nn \\
&=& \frac{2\, \hbar\, \sqrt{\beta}}{\pi\, \left(\lambda-\lambda^{'}\right)} \sin\left(\frac{(\lambda-\lambda^{'})}{2\, \hbar\, \sqrt{\beta}} \pi\right).
\eea

As function of $\lambda - \lambda^{'}$ normalized to $\hbar\, \sqrt{\beta}$, $\langle \phi_{\lambda^{'}} | \phi_{\lambda} \rangle$ was studied \cite{16}. It was found that the standard position eigenstates are no longer orthogonal, because the formal position eigenvectors $|\phi_{\lambda} \rangle$ are not physical states, i.e. not part of the domain of $\texttt{p}$. In other words, they have infinite uncertainty in momentum and  in particular infinite energy $\left\langle \phi_{\lambda} \left| \textbf{p}^{2}/(2\, m) \right| \phi_{\lambda} \right\rangle = \texttt{divergent}$  \cite{16}.

\subsubsection{Maximal localization states} 
\label{Localization}

The maximum localization around position states $|\phi_{\zeta}^{ml}\rangle$, $\left\langle \phi_{\zeta} ^{ml}\left| \hat{X} \right|\phi_{\zeta} ^{ml} \right\rangle = \zeta$, and $\Delta x_{min} = \Delta x_{0}$ depends on $\langle p \rangle$, which satisfy the inequality \cite{16}
\bea  
\left|\left|\left((x-\langle x \rangle)+ (p-\langle p \rangle )\frac{\left[x, p \right]}{2 (\Delta p)^{2}}\right) | \phi \rangle \right|\right| \geq 0. \label{equaa} 
\eea 
This implies 
\bea
\Delta x\, \Delta p &\geq & \frac{1}{2}\left| \langle \left[x, p \right]\rangle \right|. 
\eea

For first-order GUP parameter, we can use the approximate relation \cite{16,Scardigli,Das1,Das}
\bea 
\left|\langle \left[x,\, p \right] \rangle\right| &\approx & i\, \hbar\, \left(1+\beta (\Delta p)^{2} +\beta\, \langle p \rangle ^{2} \right).
\eea
In the momentum space and from Eqs. (\ref{eq:app}) and (\ref{eq:app1}), this gives the differential equation \cite{16}
\bea 
\left\{\left[i\hbar(1+\beta p^{2} \right)\partial_{p}  - \langle x \rangle] + i\, \hbar\, \frac{\left(1+\beta (\Delta p)^{2} +\beta \langle p \rangle ^{2} \right)}{2 (\Delta p)^{2} } (p-\langle p \rangle )\right\} \phi (p) &\approx & 0, \nn
\eea
which can be solved  as
\bea
\phi (p) &\approx & C (1+\beta p^{2})^ \frac{-\left(1+\beta (\Delta p)^{2} +\beta \langle p \rangle ^{2} \right)}{4\, \beta\, (\Delta p)^{2}} \nn \\
&& \exp \left[\left(\frac{\langle x \rangle}{i\, \hbar\, \sqrt{\beta}} - \frac{\left(1+\beta\, (\Delta p)^{2} +\beta\, \langle p \rangle ^{2} \right) \langle p \rangle}{2\, \sqrt{\beta} (\Delta p)^{2}}\right)\; \tan^{-1}(\sqrt{\beta}\, p)\right].
\eea

At $\langle p \rangle =0$ and critical momentum uncertainty $(\Delta p)^{2}=1/\beta$, the absolutely maximal localization reads   \cite{16}
\bea
\phi ^{ml}_{\zeta} (p) &\approx & C\, (1+\beta\, p^{2})^{-\frac{1}{2}}\; \exp\left(- i\, \frac{\langle x \rangle \tan^{-1}(\sqrt{\beta} p)}{ \hbar\, \sqrt{\beta}}\right).
\eea  
The momentum space wavefunctions $|\phi_{\zeta} ^{ml} \rangle$ of a maximum localization around $\zeta$ reads
\bea
\phi ^{ml}_{\zeta} (p) &=& \sqrt{\frac{2 \sqrt{\beta}}{\pi}} \left(1+\beta p^{2}\right)^{-\frac{1}{2}} \; \exp\left(- i \frac{\zeta \tan^{-1}(\sqrt{\beta} p)}{ \hbar \sqrt{\beta}}\right).
\eea
These states generalize the plane waves in the momentum-space and describe maximal localization in the ordinary QM. This leads to proper physical states with finite energy \cite{16}
\bea 
\left\langle \phi_{\zeta} ^{ml}\left| \frac{\hat{\textbf{P}}^{2}}{2 m} \right|\phi_{\zeta} ^{ml} \right\rangle &=& \frac{2 \sqrt{\beta}}{\pi} \int_{-\infty}^{\infty} \frac{dp}{(1+\beta p^2)^{2}} \frac{p^{2}}{2m}= \frac{1}{2 m \beta}.
\eea

\subsubsection{Transformation to quasiposition wavefunctions}

Through projecting arbitrary states on maximally localized states,  the probability amplitude for the particle being maximally localized around a position can be obtained. For quasiposition wavefunction $\phi(\zeta)=\langle \phi^{ml} _{\zeta} | \phi \rangle$ \cite{16},  where in the limit $\beta \rightarrow 0$, the ordinary position wave function $\phi(\zeta)=\langle\zeta|\phi\rangle$. The quasiposition wavefunction of a momentum eigenstate $\phi_{\widetilde p} (P) = \delta (p-\widetilde p)$ with energy $E = {\widetilde p}^{2}/2\, m$ is characterized as a plane wave. The transformation of the wavefunction in momentum representation into its counterpart quasiposition wavefunction is given as \cite{16}
\bea 
\phi (\zeta) &=& \sqrt{\frac{2\, \sqrt{\beta}}{\pi}} \int_{-\infty}^{\infty} \frac{d p}{(1+\beta\, p^2)^{\frac{3}{2}}}\; \exp\left[i\, \frac{\zeta\, \tan^{-1}(\sqrt{\beta}\, p)}{ \hbar\, \sqrt{\beta}}\right]\, \phi(p). \label{ddd}
\eea
In terms of modified dispersion relation, the wavelength is given as  \cite{16}
\bea \label{quiB}
\lambda (E) =\frac{2\, \pi\, \hbar\, \sqrt{\beta}}{\tan^{-1}\sqrt{2\, m\, \beta\, E}}. 
\eea  
In absence of GUP, we get $\lambda_{0} = 4\, \hbar\, \sqrt{\beta}$, no wavelength components is allowed which is smaller than $\lambda_{0}$.
Furthermore, no arbitrarily fine ripples are possible, because the energy of short wavelength diverges when the wavelength approaches $\lambda_{0}$  
\bea \label{qui}
 E(\lambda) &=& \frac{1}{2 m \beta}\, \left(\tan \frac{2 \pi \hbar \sqrt{\beta}}{\lambda} \right)^{2}. 
\eea

The dependence of $\lambda(E)$  on $m\, E$ has been studied in ordinary QM and GUP approach at $\beta=0.2$ \cite{16}. It is obvious that Eq. (\ref{quiB}) is bounded from below and thus a nonzero minimal wavelength is likely. While the transformation, Eq. (\ref{ddd}), is Fourier type, that of a quasiposition wavefunction into a momentum-space wavefunction reads 
\bea
\phi(p) &=& \frac{1}{8\, \pi\,  \sqrt{\beta}\, \hbar} \int_{-\infty}^{\infty} \frac{d \zeta}{(1+\beta\, p^2)^{-1/2}}\, \exp\left[-i\, \frac{\zeta\, \tan^{-1}(\sqrt{\beta}\, p)}{\hbar\, \sqrt{\beta}}\right]\;  \phi(\zeta).
\eea

\section{Minimal length uncertainty: maximal momentum}
 \label{sec:max}

\subsection{Momentum modification}

Based on DSR, the GUP approach suggests modifications in the commutators \cite{12}
\bea 
\left[x_{i}\,, p_{j}\right] &=& i\, \hbar\, \left(\delta_{ij}\, (1+\beta p^{2})+2 \beta p_{i}p_{j} \right)  =  i\, \hbar\, \left[(1- l_{pl} |\textbf{p}|) \delta_{i j}+ l_{pl}^2  p_{i} p_{j} \right]. \label{dsr2} \\
\left[x_{i} ,p_{j}\right] &=& i\, \hbar\, \left[\delta_{i j} +\alpha_{1}\, p\, \delta_{i j}+ \alpha_{2} \frac{p_{i}\, p_{j}}{p} + \beta_{1}\, p^{2}\, \delta_{i j} + \beta_{2}\, p_{i}\, p_{j} \right].
\eea 
Then from Jacobi identity, it follows that
\bea
- \left[[x_{i}, x_{j}], p_{k} \right] = \left[[x_{j}, p_{k}], x_{i}\right] + \left[[p_{k}, x_{i}], x_{j}\right] & = & 0, \label{uuu2}\\
\left[\left(\frac{\alpha_{1} - \alpha_{2}}{p}\right) + \left(\alpha_{1}^{2} + 2\beta_{1} - \beta_{2}\right) \right] \Delta_{j k i } &=& 0,
\eea
where $\Delta _{jki} = p_{i} \delta_{jk} - p_{j} \delta_{ik}$. It was assumed that $\alpha_{1}=\alpha_{2}=-\alpha$, where the negative sign appearing in Eq. (\ref{uuu2}) or Eq. (\ref{dsr2}). At $\alpha>0$, then $\alpha_{1}^{2}+2\,\beta_{1}-\beta_{2}=0$ has the roots $\beta_{1}=\alpha^{2}$ and $\beta_{2}=3 \alpha ^{2}$ with $\alpha^{2}=\beta$. The resulting commutators are consistent with the string theory, black holes physics and DSR
\bea
\left[x_{i}, p_{j} \right]=i \hbar \left[ \delta _{i j} -\alpha \left( p \delta _{i j} +\frac{p_{i} p_{j}}{p} \right)+\alpha ^{2} \left( p^{2} \delta _{ij} +3 p_{i} p_{j} \right)\right]. \label{ali1}
\eea
By Jacobi identity, 
\bea
\left[x_{i} ,x_{j}\right]=\left[p_{i} ,p_{j}\right] &=& 0, \label{ali2}
\eea
where $\alpha = \alpha_{0} \, \ell_{pl}/\hbar= \alpha_{0}/ (M_{pl}\, c)$ and the Planck length  $\ell _{pl} \approx 10^{-35}~$m and energy $\epsilon_{pl} = M_{pl} c^{2} \approx 10^{19}~$GeV.

At $1$-dimension, this GUP approach was formulated as \cite{advplb,Das:2010zf}
\bea 
\Delta x\, \Delta p &\geq & \frac{\hbar}{2} \left(1-2\, \alpha\, \langle p \rangle +4\, \alpha^{2} \langle p^{2} \rangle \right). \label{ali3}
\eea
It is obvious that $(\Delta p)^{2}=\langle p^{2}\rangle -\langle p\rangle^{2}$ and therefore 
\bea 
\Delta x\, \Delta p &\geq &\frac{\hbar}{2} \left[1+\left(\frac{\alpha}{\sqrt{\langle p^{2} \rangle}}+4\, \alpha^{2} \right)(\Delta p)^{2} + 4\, \alpha^{2}\, \langle p \rangle^{2} - 2\, \alpha\, \sqrt{\langle p^{2} \rangle} \right].
\eea
The commutators and inequalities similar to the ones given in Eqs. (\ref{ali1})  and (\ref{ali3}) have been proposed and derived in Ref. \cite{advplb,Das:2010zf}. This implies a minimum measurable length and a maximum measurable momentum, simultaneously
\bea
\Delta x &\geq & (\Delta x )_{min} \approx \alpha\, \hbar \approx \alpha_{0}\, \ell_{pl}, \label{app1} \\
\Delta p &\le &  (\Delta p )_{max} \approx \frac{1}{\alpha} \approx \frac{M_{pl}\, c}{\alpha _{0}}, \label{app2}
\eea
and defines 
\bea 
X_{i} &=& x_{0 i}, \label{ali555}\\ 
P_{j} &=& p_{0 j} (1 - \alpha\, p_{0} +2\, \alpha^{2}\, p_{0}^{2}). \label{ali222}
\eea 
We note that $p_{0}^{2}=\sum_{j}^3\,p_{0 j}\,p_{0 j}$ satisfies the canonical commutation relations $\left[x_{0 i}, p_{0 j}\right] = i \hbar \delta_{i j}$ and $p_{0j}$ is defined as the momentum at low-energy scale, which is represented by  $p_{0 j}=-\,i\,\hbar\;\de/\de x_{0 j}$, while $p_{j}$ is considered as the momentum at high-energy scale. It is assumed that the dimensionless parameter $\alpha_{0}$ has value very close to unity. In this case, the $\alpha$-dependent terms are important only when the energies (momenta) are comparable to the Planck energy (momentum), and the lengths are comparable to the Planck length.


Regardless the wide range of applications in different physical systems, crucial difficulties are listed out \cite{pedrama,pedramb}:
\begin{itemize}
\item It contains linear and quadratic terms of momenta with a minimum measurable length and a maximum measurable momentum. 
\item It was claimed that when the energy becomes close the Planck limit, there should be a modification in Eq. (\ref{ali222}) and this should ensure commutators of space, Eq. (\ref{ali2}), as the canonical system, which can predict the measurable length and a maximum measurable momentum, simultaneously. 
\item it is a perturbative approach. Therefore, it is only valid for small values of the GUP parameter $\alpha$,
\item it can not approach the non-commutative geometry, see Eq. (\ref{ali2}),
\item it suggests a minimal length uncertainty which can be interpreted as the minimal length. The maximal momentum uncertainty differs from the idea of the maximal momentum which is required in DSR theories, where the maximal momentum given in uncertainty not on the value of the observed momentum, see Eq. (\ref{app2}), 
\item it suggests momentum modification given in Eq. (\ref{ali222}), but does not achieve the commutator relation of the momentum space $[p_{i},p_{j}] \ne 0$, 
\item its minimal length uncertainty with maximal momentum results in uncertainty instead of  maximum observed momentum, see Eqs. (\ref{app1}) and (\ref{app2}), and
\item the introduction of the minimal length (non-varnishing value) allows the study for  the Hilbert space representation corresponding to the momentum wavefunction $\psi(p)$. 
\end{itemize}

\subsection{Hilbert space representation}

In the first term of Eq. (\ref{ali3}), which is related to the momentum (refers to maximal momentum), various differences between the Hilbert space representation and the work of KMM \cite{16} can be originated. Assuming that the minimal observable length has a non-vanishing uncertainty, one should construct a new Hilbert space representation, which is compatible with the commutation relation accompanied with the GUP approach 
\bea
[x_{i}\,, p_{j}] &=& i\, \hbar\,  \delta_{i j}\, \left(1-\alpha\,  p  +2\, \alpha^{2}\,  \vec{p}^{2} \right).
\eea
But, when neglecting the minimal momentum uncertainty, there would still exist a continuous momentum space representation. This means that various physical applications of the minimal length by implementing convenient representation of the commutation relations on momentum-space wavefunctions can be explored \cite{amir}
\bea 
\textbf{X}_{i}\, \phi (p) &=& x_{0 i}(1 -\alpha p_{0} +2\, \alpha^{2}\, \vec{p_{0}}^{2})\, \phi (p), \label{amm1}\\
\textbf{P}_{j}\, \phi (p) &=&  p_{0 j}\, \phi (p), \label{amm2}
\eea
where $p_{0}^{2} = \sum_{j}^3\, p_{0 j}\, p_{0 j}$ satisfying the canonical commutation relations $\left[x_{0i },\, p_{0 j}\right] = i\, \hbar\, \delta_{i j}$ and $p_{0 j}$ is defined as the momentum at low-energy scale which is represented by  $x_{0 i}=i\, \hbar\, \de_{p_{i}}$. 

These commutation relations imply a nonzero minimal uncertainty in each position coordinate (in ordinary QM,  $\left[p_{i},\, p_{j}\right]=0$). Then, it is straightforward to show that 
\bea
\left[x_{i},\, x_{j}\right] &=& i\, \hbar\, \alpha\, \left(4\, \alpha - \frac{1}{P}\right)\; \left(\textbf{P}_{i} \textbf{X}_{j} - \textbf{P}_{j}\, \textbf{X}_{i} \right).
\eea
In light of this, one should be worry about the divergence in the KMM formalism \cite{16}, at vanishing momentum. Therefore, ''Singularity'' is likely, because the derivative diverges at $p=0$. However The commutation relations do not violate the rotational symmetry, the main difference with the ordinary QM appears in the relation
\bea
\left[x_{i},\, x_{j} \right] &=& i\, \hbar\, \alpha\, \left(4\, \alpha - \frac{1}{P}\right)\, \left(1 - \alpha\, p_{0} +2 \alpha^{2}\, \vec{p_{0}}^{2}\right)\; L_{i j},  \label{eq:xixj}
\eea
$L_{i j} = (\textbf{X}_{i}\,  \textbf{P}_{j} -\textbf{X}_{j}\, \textbf{P}_{i})/(1 -\alpha\, p_{0} +2\, \alpha^{2}\, \vec{p_{0}}^{2})$ are the rotation generators ($\textbf{X}$ and $\textbf{P}$ are position and momentum operators, respectively). 
The action on a momentum-space wave function 
\bea
L_{i j} \phi(p) &=& -\, i\, \hbar\,  \left( p_{i}\, \partial_{p_{j}} - p_{j}\, \partial_{p_{i}}\right)\, \phi(p).
\eea

In the original KMM formalism \cite{16}, the $1/P$-term, which represents trace of effect of the maximal momentum, does not exist. The previous equation (\ref{eq:xixj}) express the noncommutative nature of the spacetime manifold at the Planck scale. 
\begin{itemize}
\item The existence of an upper bound of momentum fits well with DSR. In this representation, the scalar product should be modified due to the presence of the additional factor $(1-\alpha\, p_{0} +2 \, \alpha^{2}\, \vec{p_{0}}^{2})$ and the maximal momentum. 
\item The integrals are calculated between $-p_{pl}$ and  $+p_{pl}$, Planck momenta. This differs from the integration region in the KMM formalism \cite{16,amir} and thus implies the existence of a maximal Planck momentum, $p_{pl} \equiv M_{pl}\, c$.
\bea
\langle \phi | \psi \rangle &=& \int_{- p_{pl}}^{+p_{pl}}\, \frac{\phi^{*} (p)\; \psi (p)}{(1 -\alpha\, p_{0} +2\, \alpha^{2} p_{0}^{2} )}\, d p.
\eea
\item  Accordingly, the identity operator  is given as \cite{amir}
\be 
\int_{- p_{pl}}^{+p_{pl}} \frac{|p \rangle \langle p|}{\left(1 -\alpha\, p_{0} +2\, \alpha ^{2}\, p_{0}^{2}\right)}\, d p = 1,
\ee
and the scalar product of the momentum eigenstates should be changed to 
\be 
\langle p | p^{'} \rangle = \left(1 -\alpha\, p_{0} +2\, \alpha^{2}\, p_{0}^{2} \right)\, \delta \left(p - p^{'} \right).
\ee
\end{itemize}

\subsubsection{Eigenstates of position operator in momentum space}

It was proposed  \cite{16,amir} that the position operator acting on the momentum-space eigenstates 
$\textbf{X}\,.\,\phi_{\xi} (p) = \xi\, \phi_{\xi} (p)$, where $\phi_{\xi}(p)=\langle \xi | p \rangle$ is the position eigenstate with $|\xi \rangle$ being an arbitrary state
\bea
i\, \hbar\, \left(1 -\alpha p_{0} + 2 \alpha ^{2} p_{0} ^{2} \right)\, \de_{p}\,  \phi_{\xi}(p) &=& \xi\, \phi_{\xi}(p).
\eea
By solving this differential equation, the formal position eigenvectors can be derived \cite{amir}
\bea 
\phi_{\xi} (p) &=& C\, \exp \left[-i\, \frac{2\, \xi}{\alpha\, \hbar\, \sqrt{7}} \left(\tan^{-1}\, \frac{1}{\sqrt{7}}+\tan^{-1}\, \frac{4\, \alpha\, p -1}{\sqrt{7}} \right) \right].
\eea
The formal position eigenvectors in the momentum-space can be deduced when the factor  $C$ is extracted and  normalized condition is applied \cite{amir}
\bea
\phi_{\xi} (p) &=&\sqrt{\frac{\alpha\, \sqrt{7}}{2}} \left[\tan^{-1}\, \left(\frac{4\, \alpha\, p_{pl} -1}{\sqrt{7}} \right) +\tan^{-1} \left(\frac{4\, \alpha\, p_{pl} +1}{\sqrt{7}} \right)\right]^{-\frac{1}{2}}\nonumber \\ 
&& \hspace*{7mm}\exp \left[-i\, \frac{2\, \xi}{\alpha\, \hbar\, \sqrt{7}} \left(\tan^{-1} \left(\frac{1}{\sqrt{7}}\right)+\tan^{-1}\left(\frac{4\, \alpha\, p -1}{\sqrt{7}}\right) \right) \right]. \label{eq:gmse}
\eea

The previous expression (\ref{eq:gmse}) represents generalized momentum-space eigenstate of the position operator in presence of both minimal length and maximal momentum. The scalar product of the formal position eigenstates \cite{amir}
\bea
\langle \phi_{\xi^{'}} |\phi_{\xi} \rangle &=& \int_{- p_{pl}}^{+p_{pl}} \frac{dp}{(1 -\alpha\, p_{0} +2\, \alpha^{2}\, p_{0}^{2})} \phi_{\xi^{'}}^{*}(p)\, \phi_{\xi}(p), \nn\\
 &=&  {\frac{\alpha\, \sqrt{7}}{2}}\, \rho_{0}\, \exp\left[-i\, \frac{2\, \left(\xi - \xi^{'}\right)}{\alpha\, \hbar\, \sqrt{7}}\, \tan^{-1}\left(\frac{1}{\sqrt{7}}\right)\right] \nonumber \\ 
 & &  \int_{- p_{pl}}^{+p_{pl}}\, \frac{\exp\left[{ -i\, \frac{2\, \left(\xi - \xi^{'}\right)}{\alpha\, \hbar\, \sqrt{7}}}\,  \tan^{-1}\left(\frac{4\, \alpha\, p -1}{\sqrt{7}}\right)\right]} {(1 -\alpha\, p_{0} +2\, \alpha^{2}\, p_{0}^{2} )}\, dp, \hspace*{10mm}
\eea
where $\rho_{0}=\left[\tan^{-1} \left(\frac{4\, \alpha\, p_{pl} -1}{\sqrt{7}} \right) +\tan^{-1}\left(\frac{4\, \alpha\, p_{pl} +1}{\sqrt{7}} \right)\right]^{-1}$ and therefore,
\bea
\langle \phi_{\xi^{'}} |\phi_{\xi} \rangle &=&  \Omega\, \left[ \exp\left\{-i\, \left[\frac{2 (\xi - \xi ^{'})}{\alpha \hbar \sqrt{7}}\, \tan^{-1}\left(\frac{4 \alpha p_{pl} -1}{\sqrt{7}}\right) -\frac{\pi}{2} \right]\right\}\right. \nn \\
 & - & \left.  \exp\left\{ i\, \left[ \frac{2 (\xi - \xi ^{'})}{\alpha \hbar \sqrt{7}}\, \tan^{-1}\left(\frac{4 \alpha p_{pl} +1}{\sqrt{7}}\right)+\frac{\pi}{2} \right]\right\}\right], 
\eea
with $\Omega = \frac{\rho_{0}\, \hbar\, \alpha\, \sqrt{7}}{2\, \left(\xi - \xi ^{'}\right)}\; \exp\left[-i\, \frac{2\, \left(\xi - \xi ^{'}\right)}{\alpha\, \hbar\, \sqrt{7}}\; \tan^{-1}\, \left(\frac{1}{\sqrt{7}}\right)\right]$.

For the formal position eigenvectors, the expectation value of the energy reads 
\bea
\left\langle \phi_{\xi} \left| \textbf{p}^{2}/(2\, m) \right| \phi_{\xi} \right\rangle &=& \int_{- p_{pl}}^{+p_{pl}} \phi_{\xi^{'}}^{*}\, p\, \frac{p^{2}}{2\, m} \frac{\phi_{\xi}(p)}{1 -\alpha\, p +2\, \alpha^{2}\, p^{2}}\, d p \\
& =& \frac{\alpha\, \sqrt{7}\, \rho_{0}}{4\, m} \int_{- p_{pl}}^{+p_{pl}} \phi_{\xi ^{'}}^{*}\, p\, \frac{p^{2}}{2 m} \frac{\phi_{\xi} (p)}{1 -\alpha p +2 \alpha ^{2} p ^{2}} \, dp,
\eea 
and therefore  \cite{amir}
\bea
\left\langle \frac{\textbf{p}^{2}}{2\, m} \right\rangle &=& \left[\frac{\sqrt{7} \rho P_{pl}}{4\, m} + \frac{\sqrt{7}\, \rho}{32\, m\, \alpha}\; \ln \left(\frac{1 -\alpha\, p_{pl} +2\, \alpha^{2}\, p_{pl}^{2}}{1 +\alpha\, p_{pl} +2\, \alpha^{2}\, p_{pl}^{2}} \right) -\frac{3}{16\, m\, \alpha} \right].
\eea

About this GUP approach, few remarks are on order now
\begin{itemize}
\item As shown in previous sections, the energy spectrum is not divergent as the one related to the framework of KMM GUP-approach \cite{16}, especially in the presence of both minimal length and maximal momentum,
\item but, it turns out also that the expectation values of the energy as calculated by the GUP approach \cite{advplb,Das:2010zf,afa2} are no longer divergent \cite{amir}. 
\item It should be highlighted that the expectation values of energy are not lying within the domain of $P$, which physically means that they have infinite momentum uncertainty.  
\end{itemize}

\subsubsection{Maximal localization states}

In order to calculate the states $|\phi_{\zeta}^{ml} \rangle$ of the maximum localization around the position $\zeta$, it should be assumed that $\left\langle \phi_{\zeta}^{ml} \left| \hat{X} \right| \phi_{\zeta}^{ml} \right\rangle = \zeta$   \cite{16}.  As in section \ref{Localization} and by using Eqs. (\ref{amm1}) and (\ref{amm2}) and the differential equation in momentum space, Eq. (\ref{equaa}), then  
\bea 
\left\{\left[i\hbar(1-\alpha p+2\alpha ^{2} p^{2} ) \partial_{p} - \langle \textbf{X} \rangle\right] + i\, \hbar\, \frac{1+2\alpha ^{2} (\Delta p)^{2} + \alpha^{2} \langle p \rangle^{2} -\alpha\, \langle p \rangle}{2 (\Delta p)^{2} } (p-\langle p \rangle)\right\} \phi (p) &\approx & 0 \nn
\eea
When taking into account that $\langle \textbf{X} \rangle=\zeta$, $\langle p \rangle=0$ and $\Delta p=\alpha/2$, the minimal position uncertainty can be deduced from the solution of this differential equation, which are correspondent to the states of absolutely maximal localization and critical momentum uncertainty.  By normalization where the Planck momentum is of the order of magnitude as that of $P_{pl}=\alpha/2$, then $\eta=(4 \alpha p_{pl} - 1)/\sqrt{7}=3/\sqrt{7}$. Therefore, the momentum-space wavefunctions $\phi_{\zeta}^{ml} (p)$ of states, which are maximally localized around $\langle \textbf{X} \rangle = \zeta$ \cite{amir}
\bea 
\phi_{\zeta}^{ml} (p) &=& \frac{\sqrt{6 \alpha}\left[\sqrt{8}\, e^{\eta \tan^{-1}(\eta)} - e^{-\eta \tan^{-1}\left(\frac{\eta}{3}\right)}\right]^{-\frac{1}{2}}}{(1 +\alpha p +2 \alpha^{2} p^{2} )^{\frac{3}{4}}} \nn \\
&& e^{\frac{-\eta}{2} \tan^{-1}\left(\frac{4 \alpha p - 1 }{\sqrt{7}}\right)}\; e^{{-i \frac{2 \zeta}{\alpha \hbar \sqrt{7}} } \left(\tan^{-1}\left(\frac{\eta}{3}\right) + \tan^{-1}\left(\frac{4 \alpha p - 1 }{\sqrt{7}}\right) \right)}. \hspace*{1cm}
\eea

It is apparent that the difference between this result and the one which was obtained in framework of  KMM GUP  \cite{16} is due the presence of first-order  momentum, Eq. (\ref{amm1}), which implies the existence of a maximal momentum. The  maximal localization states are now the proper physical states of the finite energy \cite{amir}
\bea 
\left\langle \phi_{\zeta}^{ml}\left| \frac{\hat{\textbf{P}}^{2}}{2\, m} \right|\phi_{\zeta}^{ml} \right\rangle &=& \frac{2\, \sqrt{\beta}}{\pi} \int_{- p_{pl}}^{+p_{pl}} \frac{\phi_{\zeta}^{ml*}(p)\, \frac{p^{2}}{2\, m}\, \phi_{\zeta}^{ml}(p)}{(1 -\alpha p +2 \alpha^{2} p^{2} )}\, d p. 
\eea
This can be approximated as $\approx (32\, m\, \alpha^{2})^{-1}$.

\subsubsection{Quasiposition wavefunction transformation}

When projecting arbitrary states to maximally localized states, the probability amplitude for the particle can be deduced. This is maximally localized around a concrete position \cite{16,amir}. The transformation of a state in momentum wavefunction representation into its quasiposition wavefunction looks as \cite{amir} 
\bea
\phi (\zeta) &=& A \int_{- p_{pl}}^{+p_{pl}} \frac{\exp\left[\frac{-\eta}{2} \tan^{-1}\left(\frac{4\, \alpha\, p - 1 }{\sqrt{7}}\right)\right]}{\left[1 +\alpha\, p +2 \alpha^{2}\left(p ^{2}\right)\right]^{\frac{7}{4}}}\; \exp(i\, H\, \zeta),
\eea
where $A = \sqrt{6 \alpha}\; \left[\sqrt{8}\, \exp(\eta \tan^{-1}(\eta))  - \exp\left(-\eta \tan^{-1}(\eta/3)\right)\right]^{-\frac{1}{2}}$ and 
$H = 2/(\alpha \hbar \sqrt{7})\, \left[\tan^{-1}(\eta/3) + \tan^{-1}((4\, \alpha\, p - 1)/\sqrt{7}) \right]$ are modified wavenumbers.
Then, the modified wavelength in quasiposition wavefunction representation for the physical states reads $\lambda (p) = \pi\, \alpha\, \hbar\, \sqrt{7}/[\tan^{-1}\left(\frac{\eta}{3}\right) + \tan^{-1}\left((4\, \alpha\, p - 1)/\sqrt{7}\right) ]$. 
Because $\alpha$ is non-vanishing and $p$ is limited to the Planck momentum, there should be no wavelength smaller than $\lambda_{0} = \lambda (p_{pl}) = (\pi\, \alpha\, \hbar\, \sqrt{7})/(\tan^{-1}(\eta/3) + \tan^{-1}[(4\, \alpha\, p_{pl} - 1)/\sqrt{7}])$.
By implementing the relation between energy and momentum, for instance through $E=p^{2}/2 m$, we get the energy
\bea
E(\lambda) &=& \frac{2}{m \alpha^{2}} \left( \frac{\tan\left(\frac{\hbar \pi \alpha \sqrt{7}}{\lambda }\right)}{ \tan\left(\frac{\hbar  \pi \alpha \sqrt{7} }{\lambda }\right) + \sqrt{7}} \right)^{2},
\eea
and $E(\lambda_{0})=(P_{pl}^{2})/(2\, m)$ which apparently agrees well with ordinary QM. 

In this approach, 
\begin{itemize}
\item all these expressions do not diverge, 
\item they are important that they are distinguishable from the KMM \cite{16}, where the quasiposition wavefunctions in contrast to the ordinary QM {\it ripples}, because the energy of the short wavelength modes is divergent and
\item similar to the ordinary QM, those wavefunctions have ordinary {\it fine ripples}, because no longer divergence in the energy at $\lambda_{0}$ takes place. 
\end{itemize}
These are important results from this new GUP approach, especially the one, which guarantees both minimal length and maximal momentum.

\section{Higher-order GUP} 
\label{other}

Other GUP approaches propose higher-order modifications and solve some of the physical constraints/problems appeared when applying either linear or quadratic GUP approaches. One alternative approach gives predictions for the minimal length uncertainty, section \ref{sec:1HO}.  Second one foresees maximum momentum besides the minimal length uncertainty, section \ref{sec:2HO}. An extensive comparison between three GUP approaches is elaborated in section \ref{sec:cpmprsn}.

\subsection{Minimal length uncertainty}
\label{sec:1HO}

Nouicer suggested a higher-order GUP approach \cite{Nouicer}. To the leading order, this agrees well with the GUP given in Eq. (\ref{eg}), predicts a minimal length uncertainty and assures Heisenberg algebra, $\left[x,\, p\right] = i\, \hbar\, \exp\left(\beta\, p^{2}\right)$.
Apparently, this algebraic basis can be fulfilled from the representation of position and momentum operators 
\bea
X\, \psi(p) &=& i\, \hbar\, \exp\left(\beta\, p^{2}\right) \de_{p}\, \psi(p), \\
P\, \psi(p) &=& p\, \psi(p), 
\eea
which are symmetric and imply modified completeness relation 
\bea
\langle \phi | \psi \rangle &=& \int_{-\infty}^{\infty} dp\, \exp \left(- \beta\, p^{2}\right)\, \phi ^{*}(p)\,   \psi (p).
\eea
The scalar product of the momentum eigenstates changes to $\langle p | p^{'} \rangle = \exp\left(\beta\, p^{2}\right)\, \delta (p - p^{'})$.
Also, the absolutely smallest position uncertainty is given as
\bea
(\Delta\, x)_{min} &=& \sqrt{\frac{e}{2}}\, \hbar\, \sqrt{\beta}.
\eea

\subsection{Minimal length  and maximal momentum uncertainty}
\label{sec:2HO}

Another  higher-order GUP approach was proposed in Ref. \cite{pedrama,pedramb}, assuming $n$-dimensions and implying both minimal length uncertainty and maximal observable momentum, 
\bea
[X_{i},\, P_{j}] &=& \frac{i\, \hbar}{1 - \beta\, p^{2}}\, \delta _{i j},
\eea 
where $p^{2} = \sum_{j}^3\, p_{j}\, p_{j}$. If the components of the momentum operator are assumed to commutate, $[P_{i},\, P_{j}] = 0$.
The Jacobi identity determines the commutation relations between the components of the position operator 
\bea
[X_{i},\, X_{j}] &=& \frac{2\, i\, \hbar\, \beta}{(1- \beta\, p^{2})^{2}}\, (P_{i}\, X_{j} - P_{j}\, X_{i}),
\eea
which apparently results in a non-commutative geometric generalization of the position space. In order to fulfil  these commutation relations, the position and momentum operators in the momentum space representation should  be written as
\bea 
X_{i}\, \phi (p) &=& \frac{i\, \hbar }{1- \beta\, p^{2}} \de_{p_{i}} \phi (p),\\
P_{j}\, \phi (p) &=& p\, \phi.
\eea
In $1$-dimension, the symmetricity condition of the position operator implies modified completeness relation with a domain varying from $-1/\sqrt{\beta}$  to $+1/\sqrt{\beta}$ \cite{pedrama,pedramb} 
\bea
\langle \phi | \psi \rangle &=& \int_{-1/\sqrt{\beta}}^{+1/\sqrt{\beta}} dp\; (1-\beta p^{2}) \phi ^{*} (p)  \psi (p).
\eea
Apparently, this result differs from KMM \cite{16}.

Furthermore, the scalar product of the momentum eigenstates will be changed to $\langle p | p^{'} \rangle = \delta (p - p^{'})/(1-\beta p^{2})$. Also, the particle's momentum is bounded from above, $P_{max} = 1/\sqrt{\beta}$.
The presence of an upper bound agrees with DSR \cite{12,13}. As we shall see, the physical observables such as energy and momentum are not only non-singular, but they are  also bounded from above, as well. The absolutely smallest uncertainty in position reads
\bea
(\Delta X)_{min} &=& \frac{3\, \sqrt{3}}{4}\;	\hbar\, \sqrt{\beta}.
\eea

\begin{itemize}
\item This new GUP approach \cite{pedrama,pedramb} estimates the minimal length uncertainty and the maximal observable momentum, simultaneously. 
\item It includes a quadratic term of the momentum and apparently assures non-commutative geometry. 
\item The maximal observable momentum agrees with the one estimated in DSR \cite{12,13}. 
\item If the binomial theorem is applied on this GUP approach, the GUP approach which was predicted in string theory \cite{guppapers,2}, black hole physics \cite{3,7} can be reproduced.  
\end{itemize}

On the other hand, it is worthwhile to notice that this new GUP* approach \cite{pedrama,pedramb} does not agree with the commutators relation which was predicted in DSR \cite{12,13}. The latter contains a linear term of momentum that is responsible for the existence of maximal observable momentum. 

\subsection{Comparison between higher-order GUP approaches}
\label{sec:cpmprsn}

\begin{table}[htb]
\label{tab:1}
\begin{center}
\begin{tabular}{|c||c|c|c|}
\hline 
Comparison  & KMM \cite{16} & ADV \cite{advplb,Das:2010zf} &Pedram \cite{pedrama,pedramb} \\ 
\hline  \hline
Algebra $\left[x,\, p\right]$ & $i \hbar \left(1+\beta p^{2}\right)$ & $i \hbar \left(1-\alpha p +2 \alpha ^{2} p^{2} \right)$& $\frac{i \hbar}{1-\beta p^{2}}$ \\ 
\hline 
$ (\Delta x)_{min} $ & $ \hbar \sqrt{\beta} $ & $ \hbar \alpha $ & $\frac{3\sqrt{3}}{4} \hbar \beta$   \\ 
\hline 
${(\Delta p)_{max}}$ & - & ${\frac{M_{pl} c}{\alpha _{0}}}$ & -  \\ 
\hline 
$P_{max}$ & Divergence  & $(\frac{1}{4\alpha })$  &  $(\frac{1}{\sqrt{\beta}})$  \\  
\hline 
\begin{tabular}{c}
$ \textbf{P} . \phi(p)$ \\ 
$\textbf{X} . \phi(p)$ \\ 
\end{tabular}  & \begin{tabular}{c}
$p \phi(p)$ \\ 
$i \hbar \left(1+\beta p^{2}\right) \partial_{p} \phi(p)$ \\ 
\end{tabular}  &\begin{tabular}{c}
$ p \phi(p)$ \\ 
$i \hbar \left(1-\alpha p +2 \alpha ^{2} p^{2} \right) \partial_{p} \phi(p)$ \\ 
\end{tabular} &\begin{tabular}{c}
$p \phi(p)$ \\ 
$\frac{i \hbar}{1- \beta p^{2}} \partial_{p} \phi(p)$\\  
\end{tabular}  \\
\hline 
Geometry & $\left[x_{i} ,x_{j}\right] \ne 0$ &  $\left[x_{i} ,x_{j}\right] \ne 0$  &  $\left[x_{i} ,x_{j}\right] \ne 0 $  \\ 
\hline 
 $\langle \frac{p^{2}}{2m} \rangle _{ max-localize-state}$ &$\frac{1}{2m \beta} $& $\frac{1}{32 m \alpha ^{2}}$ & $\frac{3}{2m \beta}$ \\ 
\hline 
 $(E(\lambda )$ or $\lambda (E))_{quasi-position}$  &$\frac{1}{2 m \beta} \left(\tan \frac{2 \pi \hbar \sqrt{\beta} }{\lambda} \right)^{2} $ & $\frac{2}{m \alpha ^{2}} \left( \frac{\tan(\frac{\hbar \alpha \sqrt{7} }{\lambda })\pi}{\tan(\frac{\hbar \alpha \sqrt{7} }{\lambda })\pi + \sqrt{7}} \right)^{2}$ & $\frac{2 \pi \hbar}{(1-\frac{2}{3} m \beta E ) \sqrt{2 m E}}$ \\ 
\hline 
$\lambda_{0}\;$ of~wavefuntion & $4 \hbar \sqrt{\beta}$ & $ \frac{\pi \alpha \hbar \sqrt{7}}{\left(\tan ^{-1} \frac{\eta}{3} + \tan ^{-1} \frac{4 \alpha p_{pl} - 1 }{\sqrt{7}} \right) }$&$3 \pi \hbar \sqrt{\beta} $  \\
\hline 
\end{tabular}
\caption{A comparison between the main features of the GUP approaches that were proposed by  KMM \cite{16}, ADV \cite{advplb,Das:2010zf} and Pedram \cite{pedrama,pedramb}.  }
\end{center}
\end{table}

Tab.~\ref{tab:1} summarizes an comprehensive comparison between the GUP approaches  of KMM \cite{16}, Ali, Das, Vagenas (ADV) \cite{advplb,Das:2010zf} and  Pedram \cite{pedrama,pedramb}. The minimum position uncertainty varies from $\hbar\,\alpha$ or $\hbar\,\sqrt{\beta}$ (both are equivalent) and $\sqrt{27}\, \hbar\, \alpha/4$, respectively. There is a maximum momentum uncertainty in ADV, although, it is wrongly called maximum momentum. The maximum momentum diverges in KMM, while it remains finite, $1/4\alpha$ and $1/\sqrt{\beta}$, respectively, in ADV and Pedram. The momentum operator and resulting geometry remain unchanged in all approaches. The position operator characterizes the different approaches. The maximum localised state slightly varies. The resulting energy (wavelength) related to quasiposition and wavefuction are very characteristic.

\section{Bounds on GUP parameter}

The GUP parameter $\alpha=\alpha _0/(M_{p} c)=\alpha _0 \ell _p/\hbar$. The Planck length $\ell _p\, \approx\, 10^{-35}~$m and the Planck energy $M_p c^2 \,\approx \, 10^{19}~$ GeV.  $\alpha _0$, the proportionality constant, is conjectured to be dimensionless \cite{advplb}. In natural units $c=\hbar=1$, $\alpha$ will be in GeV$^{-1}$, while in the physical units, $\alpha$ should be in GeV$^{-1}$ times $c$. The bounds on $\alpha_0$, which was summarized in Ref. \cite{afa2,AFALI2011,Das1}, should be a subject of precise astronomical observations, for instance gamma ray bursts \cite{Tawfik:2012hz}. 
\begin{itemize}
\item Other alternatives by the tunnelling current in scanning tunnelling microscope and the potential barrier problem \cite{AFALI2012}, where the energy of the electron beam is close to the Fermi level. It was found that the varying tunnelling current relative to its initial value is shifted due to the GUP effect \cite{AFALI2011,AFALI2012}, $\delta I/I_0\approx 2.7 \times 10^{-35}$ times $\alpha _{0} ^{2}\,$. 
In case of electric current density $J$ relative to the wave function $\Psi$, the current accuracy of precision measurements reaches the level of $10^{-5}$. Thus, the upper bound $\alpha_0<10^{17}$. Apparently, $\alpha$ tends to order $10^{-2}~$GeV$^{-1}$ in natural units or $10^{-2}~$GeV$^{-1}$ times $c$ in physical units. This quantum-mechanically-derived bound is consistent with the one at the electroweak scale \cite{Das1,AFALI2011,AFALI2012}. Therefore, this could signal an intermediate length scale between the electroweak and the Planck scales  \cite{Das1,AFALI2011,AFALI2012}.
\item On the other hand, for a particle with mass $m$ mass, electric charge $e$ affected by a constant magnetic field ${\vec B}=B_{\hat z}\approx10~$Tesla, vector potential ${\vec A}= B\,x \,{\hat y}$ and cyclotron frequency $\omega_c = eB/m$, the Landau energy is shifted due to the GUP effect \cite{AFALI2011,AFALI2012} by
\bea
\frac{\Delta E_{n(GUP)}}{E_n} &=& -\sqrt{8\, m}\; \alpha\;  (\hbar\, \omega _c )^{\frac{1}{2}} \,
 \left(n +\frac{1}{2}\right)^{\frac{1}{2}}  \approx  - 10^{-27}\; \alpha_0.
\eea
Thus, we conclude that if $\alpha_0\sim 1$, then $\Delta E_{n(GUP)}/E_n$ is too tiny to be measured. But with the current measurement accuracy of $1$ in $10^3$, the upper bound on $\alpha_0<10^{24}$ leads to $\alpha=10^{-5}$ in natural units or $\alpha=10^{-5}$ times $c$ in the physical units.

\item Similarly, for the Hydrogen atom with Hamiltonian $H=H_0+H_1$, where standard Hamiltonian $H_0=p_0^2/(2m) - k/r$ and the first perturbation Hamiltonian $H_1 = -\alpha\, p_0^3/m$, it can be shown that the GUP effect on the Lamb Shift \cite{AFALI2011,AFALI2012} reads
\bea
\frac{\Delta E_{n(GUP)}}{\Delta E_n} &\approx & 10^{-24}~\alpha_0.
\eea
Again, if $\alpha_0 \sim 1$,  then $\Delta E_{n(GUP)}/E_n$ is too small to be measured, while the current measurement accuracy gives $10^{12}$. Thus, we assume that  $\alpha_0>10^{-10}$.
\end{itemize}

In light of this discussion, should we assume that the dimensionless $\alpha_0$ has the order of unity in natural units, then $\alpha$ equals to the Planck length $\approx\, 10^{-35}~$m. The current experiments seem not be able to register discreteness smaller than about $10^{-3}$-th fm, $\approx\, 10^{-18}~$m \cite{AFALI2011,AFALI2012}. We conclude that the assumption that $\alpha_0\sim 1$ seems to contradict various observations  \cite{Tawfik:2012hz} and  experiments \cite{AFALI2011,AFALI2012}. Therefore, such an assumption should be relaxed to meet the accuracy of the given experiments. Accordingly, the lower bounds on $\alpha$ ranges from $10^{-10}$ to $10^{-2}~$GeV$^{-1}$. This means that $\alpha_0$ ranges between $10^9\, c$ to $10^{17}\, c$. 

\begin{figure}[htb]
\includegraphics[width=7.cm,angle=-90]{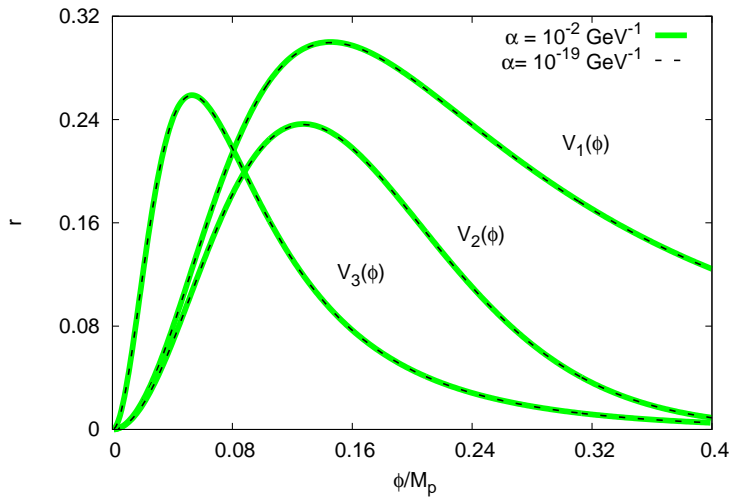}
\caption{The ratio of tonsorial-to-scalar density fluctuations, $r$, in dependence on $\phi/M_p$ calculated for the inflation potentials $V_1(\phi)$, $V_2(\phi)$ and $V_3(\phi)$. The dashed curves are evaluated at  $\alpha=10^{-2}~$GeV$^{-1}$, while the solid curves at $\alpha=10^{-19}~$GeV$^{-1}$. 
\label{spectriala}  
}
\end{figure}

\begin{figure}[htb]
\includegraphics[width=7.cm,angle=-90]{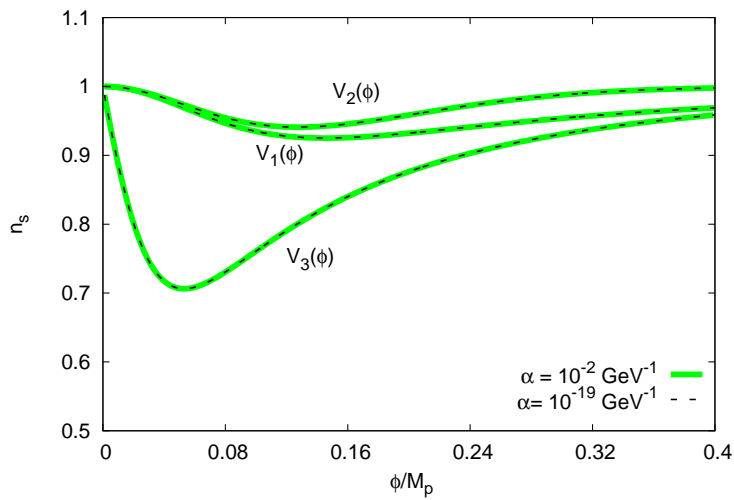}
\caption{The same as in Fig. \ref{spectriala} but for the spectral index $n_s$ vs. $\phi/M_p$. 
\label{spectrialb}  
}
\end{figure}

Fig. \ref{spectriala} shows the ratio of tonsorial to scalar density fluctuations $r$ in dependence on $\phi/M_P$. The dashed curves are evaluated at  $\alpha=10^{-2}~$GeV$^{-1}$, while the solid thick curves at $\alpha=10^{-19}~$GeV$^{-1}$. The earlier value is corresponding to $\alpha_0=10^{17}$ while the latter to $\alpha_0=1$. It is obvious that the bounds on $\alpha_0$ do no affect the ratio of tonsorial to scalar density fluctuations $r$ in dependence on $\phi/M_P$. The behavior of the tonsorial to scalar ratio is limited by the modified Friedmann equation due to GUP, where the GUP physics is related to the gravitational effect on such model at the Planck scale. The GUP parameter $\alpha$ - appearing in the modified Friedmann equation - should play an important role in bringing the value of $r$ very near to PLANCK, $r_{0.002}<0.11-0.12$  at $95\%$ confidence level.  According to Eq. (\ref{modified HH}), $\alpha$ breaks (slows) down the expansion rate. It is obvious that the parameters related to the Gaussian sections of the three curves match nearly perfectly with the results estimated by the PLANCK collaboration (compare with Fig. \ref{spectrial2}). 

Fig. \ref{spectrialb} shows the variation of the spectral index, $n_s$, with scalar field for the three inflation potentials, Eqs. (\ref{eq:mssm}), (\ref{sdual}) and (\ref{poweri}).  Again, the dashed curves are evaluated at  $\alpha=10^{-2}~$GeV$^{-1}$, while the solid thick curves at $\alpha=10^{-19}~$GeV$^{-1}$. It is obvious that the bounds on $\alpha_0$ do no affect the dependence of spectral index, $n_s$ on $\phi/M_P$.

\section{Recent cosmic inflation observations}
\label{sec:GUPRecentCosmicInflationObservations}

For a while, it was believed that the background imaging of cosmic extragalactic polarization (BICEP2) telescope at the south pole gives a possible evidence for cosmic inflation \cite{BICEP2}. Such observations wrongly misconduct an interpretation as a first direct observation for the inflation and test signatures for the quantum gravitational processes in the inflationary era, in which a primordial density and gravitational wave fluctuations are created from the quantum fluctuations \cite{Mukhanov,Bardeen}. The ratio of scalar-to-tensor fluctuation, $r$, which is a canonical measurement  of the gravitational waves  \cite{Liddle:2003,Linde:2002}, has been estimated, $r=0.2 _{-0.05}^{+0.07}$ \cite{BICEP2}. Recent PLANCK observations revealed that interstellar dust caused more than $50\%$ of the signal detected by BICEP2 \cite{planck2015a,planck2015b}. 
 
The proposed values of tensor-to-scalar ratio, $r$, require that the inflation fields are as large as the Planck scale. This idea is known as Lyth bound \cite{Lyth:96,Lyth:98,Green}, which estimates the change of the inflationary field $\Delta \phi$,
\bea
\frac{\Delta \phi}{M_p} &=& \sqrt{\frac{r}{8}} \Delta N,
\eea
Where $\Delta N$ denotes the number of e-folds corresponding to the observed scales in the CMB left the inflationary horizon. 

\subsection{Cosmic inflation models}

We apply QG approaches (GUP) in order to estimate PLANCK observations for the ratio of scalar-to-tensor fluctuation, $r$. The modified Heisenberg commutator for higher order GUP by implementing convenient representation of the commutation relations on momentum-space allows the usage of Poisson brackets between the scale factor $a$ and
momenta $p_a$ \cite{Tawfik:2014dza}
\bea 
\lbrace a\, ,\, p_a \rbrace = 1 - 2\, \alpha\, p_a.
\eea
Accordingly, the equations of motion get modifications
\bea
\label{adota}
\dot{p_a}&=& \lbrace a,\, p_a \rbrace \frac{\partial\mathcal{H}_E}{\partial a}=(1-2\alpha p_a)\,\left(\frac{2 \pi G}{3} \frac{p_a^2}{a^2} -\frac{3}{8 \pi G} \kappa + 3 a^2 \rho +a^3 \frac{d\rho}{da}\right).  \label{pdotp}
\eea
The Hamiltonian constraint reads
\bea
\mathcal{H}&=&-\frac{2\pi \, G}{3}\, \frac{p_{a}^2}{a} -\frac{3}{8\, \pi \, G}\, \kappa a\,+\, a^3 \rho \equiv 0.
\label{Hconst}
\eea
The modified Friedmann equation is
\bea
H^2 &=& 
\left(\frac{8 \pi G}{3} \rho - \frac{\kappa}{a^2}\right) \left[1\,- \, \frac{3\,\alpha \, a^2 }{ \pi G}\left(\frac{8 \pi G}{3} \rho - \frac{\kappa}{a^2}\right)^{1/2}\right].
\label{HddDo}
\eea
By taking into consideration the standard case, i.e. $\alpha$ vanishes and assuming flat Universe, i.e. $\kappa=0$, 
\bea
H^2 &=& \frac{8\, \pi \,G}{3} \rho\, \left[1-\, 3\, \alpha\, a^2  \sqrt{\frac{8}{3\, \pi \, G}}\, \rho^{1/2}\right]. \label{FR3}
\eea 

The different inflation parameters are characterized by the scalar field $\phi$ and apparently contribute to the total energy density \cite{Liddle:2003, Linde:2002}. By taking into account the cosmic background (matter and radiation) energy density, the scalar field is assumed to interact with the gravity and with itself. Under the assumption of homogeneity and isotropy of Friedmann Universe  $(\nabla \phi)^2 \ll V(\phi)\,$ \cite{Liddle:2003, Linde:2002}, also when assuming that the scalar field changes very slowly so that the acceleration would be neglected $\ddot{\phi} \ll 3\, H\, \dot{\phi}\,$ \cite{Liddle:2003, Linde:2002}, the principle condition for the expansion is where the kinetic energy is much less than the potential energy $\dot{\phi}^2 \ll  V(\phi)\,$ \cite{Liddle:2003,Linde:2002}. In order to relate the cosmological scale with the Plank mass $M_p=\left(h/c\right) \left(\Lambda/3\right)^{1/2}$, the modified Friedmann equation, Eq. (\ref{HddDo}), becomes 
\bea
H^2=\frac{4\pi}{3 M_{p}^2}\left\lbrace \left[ V(\phi)+ \frac{3 M_{p}^4}{4 \pi}\right] -3\alpha a^2 \sqrt{\frac{16\,M_{p}^2}{3 \pi}} \left[ V(\phi) + \frac{3 M_{p}^4}{4 \pi} \right]^{3/2}\right\rbrace.
\label{modified HH}
\eea

There are various cosmic  inflation models, for instance 
\begin{itemize}
\item First one is based on certain minimal supersymmetric extensions of the standard model for elementary particles~\cite{allahverdi-2006} 
\bea
V_1(\phi) = \left(\frac{m^2}{2}\right)\,\phi^2  - \left(\frac{2\sqrt{\lambda}\,m}{3}\right)\, \phi^3  + \left(\frac{\lambda}{4}\right)\,\phi^{4}, \label{eq:mssm}
\eea
which is an $\mathcal{S}$-dual inflationary potential \cite{sdual}.
\item Second one is originated in the Dirac quantization condition of the electric and magnetic charges \cite{Montonen:1977}.
\bea
V_2(\phi) &=& V_0\, \mathtt{sech} \left(\frac{\phi}{f}\right).
\label{sdual}
\eea
\item The third one is the power-law inflation model with the free parameter $d$  \cite{Starobinsky,Liddle:1993}, 
\bea
V_3(\phi)=\frac{3 M_{p}^2 d^2}{32 \pi} \left[1-\exp \left(-\frac{16\pi}{3 M_{p}^2 }^{1/2} \phi \right) \right]^2. \label{poweri}
\eea

\end{itemize}

As we shall discuss in the section that follows, the results from the first two inflation potentials disagree with the recent PLANCK observations, while the third one does not, Fig. \ref{spectrial2} and Tab. \ref{tab:1}.

\subsection{Recent PLANCK observations}

The ratio of tensor-to-scalar fluctuations are given as  \cite{Liddle:1993,Linde:1982,Liddle:1995a,Liddle:1995b}
\bea
r &=& \frac{p_t}{ p_s }\,=\,\left(\frac{\dot{\phi}}{H}\right)^2, \label{eq:r}
\eea
where the tensorial $p_t$ and scalar $p_s$ density fluctuations respectively reads \cite{Liddle:1993,Linde:1982,Liddle:1995a,Liddle:1995b}
\begin{eqnarray}
p_t &=& \left(\frac{H}{2\, \pi}\right)^{2} \left[1-\frac{H}{3\, M_p^2}\,\sin\left(\frac{6\, M_p^2}{H}\right)\right],\label{pt} \\
p_s &=& \left(\frac{H}{\dot{\phi}}\right)^2\left(\frac{H}{2\, \pi}\right)^{2}  \left[1-\frac{H}{3\, M_p^2}\,\sin\left(\frac{6\, M_p^2}{H}\right)\right]. \hspace*{10mm}
\label{ps}
\end{eqnarray}

It is obvious that the GUP effects, the $\alpha$-terms in the modified Friedmann equation, partly takes into account the QG effects. This terms reduce the denominator of the tonsorial-to-scalar density fluctuations, the Hubble parameter, $H$. Thus, the fluctuations ratio increases with decreasing $H$. 

\begin{figure}[htb]
\includegraphics[width=10cm,angle=0]{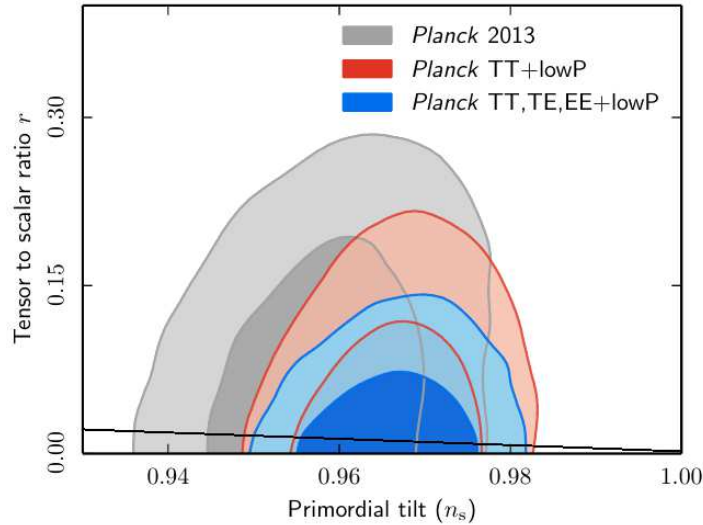}
\caption{PLANCK marginalized joint contours for $r$ vs. $n_s$ at $68\%$ and $95\%$ confidence level (red and blue contours)  compared to previous observations (gray contours), Fig. 6 in \cite{planck2015b}, and to our parametric calculations (line). 
\label{spectrial2}  
}
\end{figure}

In Fig. \ref{spectrial2}, confronts our parametric dependence of the spectral index $n_s$ and the ratio $r$ to PLANCK marginalized joint contours for $r$ vs. $n_s$ at $68\%$ and $95\%$ confidence level and previous observations \cite{planck2015b}. Both parametric quantities are functions of $\phi$, Eq. (\ref{eq:r}). We find that the recent PLANCK observations (smaller contour)  is fairly crossed by our parametric calculations from the power low inflation model, Eq.  (\ref{poweri}).  Fig. \ref{spectrial2} illustrates significant improvement (red and blue contours) with respect to previous Planck data release (gray contours). It shows no BICEP data. 

Table \ref{tab:1} summarizes scalar field inflation potential $\Delta \phi/M_{p\ell}$ and ratios of tonsorial to scalar density fluctuations $r$ within the $n_s$-region which was analysed by PLANCK.

\begin{table}[h]
\begin{tabular}{||c|c|c||}
\hline 
$\Delta \phi/M_{p\ell}$ & r & $n_s$ \\
\hline 
0.32 &       0.0128 &   0.9346 \\
\hline    
0.34  &     0.0104  &  0.9412    \\
\hline 
0.36    &    0.0084 &   0.9470    \\
\hline 
0.38     &   0.0069  &  0.9522    \\
\hline 
0.40    &  0.0056    & 0.9567    \\
\hline 
0.42    &  0.0046   & 0.9607    \\
\hline 
0.44     &   0.0038  &  0.9643   \\
\hline 
0.46   & 0.0031    & 0.9676    \\
\hline 
0.48    & 0.0026   & 0.9705 \\
\hline 
\end{tabular}
\caption{Within the $n_s$-region analysed by PLANCK, the ratios of tonsorial to scalar density fluctuations $r$  and the scalar field inflation potentials, $V_3(\phi)$, Eq.  (\ref{poweri}) are determined. \label{tab:2}}
\end{table}

So far, we conclude that depending on the inflation potential $V(\phi)$ and the scalar field $\phi$, the GUP approaches seem to explain  a considerable part of the recent PLANCK observations on the upper bound on the tensor-to-scalar ratio, $r_{0.002}<0.11$ 
at $95\%$ confidence level, when the PLANCK high-$\ell$ polarization data is included. This fits well with the upper limit according to {\it B}-mode polarization constraint, $r< 0.12$ at $95\,\%$ at confidence level, which was obtained from a joint analysis of PLANCK, BICEP2, and Keck Array data \cite{planck2015a}, Tab. \ref{tab:2}.

\section{Discussion and final remarks}
\label{sed:dis}

The Heisenberg uncertainty principle expresses one of the fundamental properties of the quantum systems. Accordingly, there should be a fundamental limit of the accuracy with which certain pairs of physical observables, such as the position and momentum, time and energy, can be measured, simultaneously. In other words, the more precisely one observable is measured, the less precise the other one can be estimated. In QM, the physical observables are described by operators acting on the Hilbert space representation of the states. Thus, the Heisenberg uncertainty principle uses operators in describing the relation between various pairs of  uncertainties.

The quantum aspects of the gravitational fields can emerge in a limit, in which the different types of interactions, like strong, weak and electromagnetism can be distinguished from each other. In string theory, the particles are conjectured to stem from fundamental strings. This fundamental scale is nothing but the string length, which is also supposed to be in order of the Planck length. The string cannot probe distances smaller than its own length.  The current researches of the quantum problems in the presence of gravitational field at very high-energy near to the Planck scale implies new physical laws and even corrections to the space-time. The quantum field theory in curved background can be normalized by introducing a minimal observable length as an effective cutoff in the ultraviolet domain. 

We have reviewed different approaches for GUP, which predict an existence of a minimal length uncertainty. The non-zero length uncertainty expresses a non-zero state in the description of the Hilbert space representation and is able to fulfil the non-commutative geometry. These should have impacts on the discreteness and the quantization of space and on the aspects related to the quantum field theory. The elicitation of the minimal length from various experiments, such as string theory, black hole physics and loop quantum gravity, imitates the quantum gravity. All of them predict corrections to the quadratic momentum in the Heisenberg algebra. Many authors represent such algebra under modification in the position operator which fits with the Hilbert space representation and takes into consideration the states of space (eigenvectors) corresponding to the energy (eigenvalue). Others represent such modified algebra by modification in the linear momentum. This is motived by momentum modification at very high energy, which is supposed to fulfil the Hilbert space representation but also approves the idea of modified dispersion relation of the energy-momentum tensor.

Doubly special relativity is conjectured to provide a GUP approach with an additional term reflecting the possibility to deduce information about the maximum measurable momentum. This new term and the one, which is related to the  minimal uncertainty on position are - in modified Heisenberg algebra - of first order of momentum. Some authors suggest a combination of all previously-proposed GUP-approaches in one concept, as anticipated in DSR and the string theory, black hole physics and Loop quantum gravity. Others prefer to revise the GUP of a minimal length in order to overcome some constraints. Another suggestion for the GUP-dependent on Feynman propagator seems to suffer from an exponential ultraviolet cutoff. All of these verify the predication of minimal length at very high energy, despite of the different physical expression or the algebraic representation of Heisenberg principle. In summary, we have different GUP-approaches with many of applications in various branches of physics.

An unambiguous experiment evidence to ensure these ideas is till missing. Some physicists prefer to deny due to their convention. Some have other objections. Here we review both points-of-views. The value of the GUP parameter remains another puzzle to be verified. For example, the principles of GR developed by Einstein are seen as solid obstacles against the interpretation of the GUP approaches, which are thought to violate the equivalence principle, for instance. In thermodynamics, the natural property of the kinetic energies is assumed to be violated under the consideration of these approaches. As a reason, the symmetries can be broken in quantum field theory. Furthermore, the value of the Keplerian orbit and the correction of the continuity equation for some fields are no longer correct. 

In the present review, we have summarized all these proposals and discussed their difficulties and applications. We aimed to elucidate some of these proposals. On the other hand, from various {\it ''gedanken''} experiments, which have been designed to measure the area of the apparent black hole horizon in QG, the uncertainty relation seems to be preformed. The modified Heisenberg algebra, which was suggested in order to investigate GUP, introduces a relation between QG and Poincare algebra. Under the effect of GUP in an $n$-dimension space, it is found that even the gravitational constant $G$ and the Newtonian law of gravity are subject of modifications. The interpretation of QM through a quantization model formulated in $8$-dimensional manifold implies the existence of an upper limit in the accelerated particles. Nevertheless, the GUP approaches given in forms of quadratic and linear terms of momenta  assume that the momenta approach some maximum values at very high energy (Planck scale).

In supporting the phenomena that uncertainty principle would be affected by QG many examples can be mentioned. In context of polymer quantization, the commutation relations are given in terms of the polymer mass scale. The standard commutation relations are conjectured to be changed or - in a better expression - generalized  at Planck  length. Such modifications are supposed to play an essential role in the quantum gravitational corrections at very high energy. Accordingly, the standard uncertainty relation of QM should be replaced by a gravitational uncertainty relation having a minimal observable length of the order of the Planck length.
On the other hand, the detectability of quantum space-time foam with gravitational wave interferometers has been addressed. The limited measurability of the smallest quantum distances has been criticized. An operative definition for the quantum distances and the elimination of the contributions from the total quantum uncertainty were given. In describing the quantum constraints on the black hole lifetime, Wigner inequalities have been applied. It was found that the black hole running time should be correspondent to the Hawking lifetime, which is to be calculated under the assumption that the black hole is a black body. Therefore, the utilization of Stefan-Boltzmann law is eligible. It is found that the Schwarzschild radius of black hole is correspondent to the constraints on the Wigner size. Furthermore, the information processing power of a black hole is estimated by the emitted Hawking radiation. 

There are several observations supporting GUP approaches and offer a valuable possibility to study the influence of the minimal length on the properties of a wide range of physical systems, especially at quantum scale. The effects of linear GUP-approach have been studied on the compact stars, the Newtonian law of gravity, the inflationary parameters and thermodynamics of the early Universe, the Lorentz invariance violation and the measurable maximum energy and minimum time interval. 
It was observed that GUP can potentially explain the small observed violations of the weak equivalence principle in neutron interferometry experiments and also predicts a modified invariant phase-space which is relevant to Lorentz transformation. It is suggested that GUP can be measured directly in Quantum Optics Lab. 

The experimental tests for Lorentz invariance become more accurate. A tiny Lorentz-violating term can be added to the conventional Lagrangian, then the experiments should be able to test the Lorentz invariance by setting an upper bound to the coefficients of this term, where the velocity of light $c$ should differ from the maximum attainable velocity of a material body. This small adjustment of the speed of light leads to modification in the energy-momentum relation and adding $\delta \textit{v}$ to the vacuum dispersion relation which could be sensitive to the type of candidates for the quantum gravity effect that has been recently considered in the particle physics literature. In additional to that, the possibility that the relation connecting energy and momentum in the special relativity may be modified at the Planck scale, because of the threshold anomalies of ultra-high energy cosmic ray (UHECR) is conventionally named as {\it Modified Dispersion Relations (MDRs)}. This can provide new and many sensitive tests for the special relativity. Accordingly, successful researches would reveal a surprising connection between the particle physics and cosmology. The speed of light not limited to that, but do many searches for the modification of the energy-momentum conservations laws of interaction such as pion photo-production by the inelastic collisions of the cosmic-ray nucleons with the cosmic microwave background and higher energy photon propagating in the intergalactic medium which can suffer inelastic impacts with photons in the Infra-Red background resulting in the production of an electron-positron pair.

The systematic study of the black hole radiation and the correction due to entropy/area relation gain the attention of theoretical physicists. For instance, there are nowadays many methods to calculate the Hawking radiation. Nevertheless, all results show that the black hole radiation is very close to the black body spectrum. This conclusion raised a very difficult question whether the information is conserved in the black hole evaporation process? The black hole information paradox has been a puzzled problem. The study of the thermodynamic properties of black holes in space-times is therefore a very relevant and original task. For instance, based
on recent observation of supernova, the cosmological constant may be positive. The possible corrections can be calculated by means of approaches to the quantum gravity. Through the comparison of the corrected results obtained from this alternative approaches, it can be shown that a suitable choice of the expansion coefficients in the modified dispersion relations leads to the same results in the GUP approach. 

The existence of minimal length and maximum momentum accuracy is preferred by various physical observations. Thought experiments have been designed to illustrate influence of the GUP approaches on the fundamental laws of physics, especially at the Planck scale. The concern about the compatibility with the equivalence principles, the universality of gravitational redshift and the free fall and  reciprocal action law should be addressed. The value of the GUP parameters remains a puzzle to be verified. Furthermore, confronting GUP approaches to further applications would elaborate essential properties. The ultimate goal would be an empirical evidence that the same is indeed quantized and its fundamental is given by the minimal length accuracy. If the current technologies would not able to implement this proposal, we are left with the empirical prove that the modifications of various physical systems can be estimated, accurately. To this destination, we should try to verify the given approaches, themselves. We believe that the compatibility with MDR would play the role of the Rosetta stone translating GUP in energy-momentum relations. The latter would have cosmological and astrophysical observations.
  
\section*{Acknowledgement}
This work is partly supported by the World Laboratory for Cosmology And Particle Physics (WLCAPP), http://wlcapp.net/. The authors are very grateful to the anonymous referee for her/his constructive comments and constructive comments, which contribute a lot to  improving the script! \\

\end{document}